\documentclass[preprint,twocolumn,10pt,a4paper,byrevtex,prb,unsortedaddress,superscriptaddress]{revtex4}

\usepackage{amssymb}
\usepackage{natbib}
\usepackage{graphicx}
\usepackage{amsmath}
\usepackage{float}
\usepackage{verbatim}
\usepackage{color}
\usepackage{bbold} 
\usepackage{hyperref}

\begin{document}
\title{Interfacial Spin-Orbit Coupling: A Platform for Superconducting Spintronics}

\author{Isidoro Mart\'inez}
\thanks{These authors contributed equally to the manuscript}
\affiliation{Departamento F\'isica de la Materia Condensada C-III, Instituto Nicol\'as Cabrera (INC) and  Condensed Matter Physics Institute (IFIMAC), Universidad Aut\'onoma de Madrid, Madrid 28049, Spain}

\author{Petra H\"ogl}
\thanks{These authors contributed equally to the manuscript}
\affiliation{Institute for Theoretical Physics, University of Regensburg, 93040 Regensburg, Germany}

\author{C\'esar Gonz\'alez-Ruano}
\thanks{These authors contributed equally to the manuscript}
\affiliation{Departamento F\'isica de la Materia Condensada C-III, Instituto Nicol\'as Cabrera (INC) and  Condensed Matter Physics Institute (IFIMAC), Universidad Aut\'onoma de Madrid, Madrid 28049, Spain}

\author{Juan Pedro Cascales}
\affiliation{Francis Bitter Magnet Laboratory, Massachusetts Institute of Technology, Cambridge, Massachusetts 02139, USA}

\author{Coriolan Tiusan}
\affiliation{Department of Physics and Chemistry, Center of Superconductivity Spintronics and Surface Science C4S, Technical University of Cluj-Napoca, Cluj-Napoca, 400114, Romania}
\affiliation{Institut Jean Lamour, Nancy Universit\`{e}, 54506 Vandoeuvre-les-Nancy Cedex, France}

\author{Yuan Lu}
\affiliation{Institut Jean Lamour, Nancy Universit\`{e}, 54506 Vandoeuvre-les-Nancy Cedex, France}

\author{Michel Hehn}
\affiliation{Institut Jean Lamour, Nancy Universit\`{e}, 54506 Vandoeuvre-les-Nancy Cedex, France}

\author{Alex Matos-Abiague}
\affiliation{Department of Physics and Astronomy, Wayne State University, Detroit, MI 48201, USA}

\author{Jaroslav Fabian}
\affiliation{Institute for Theoretical Physics, University of Regensburg, 93040 Regensburg, Germany}

\author{Igor \v{Z}uti\'c}
\email[e-mail: ]{zigor@buffalo.edu}
\affiliation{Institute for Theoretical Physics, University of Regensburg, 93040 Regensburg, Germany}
\affiliation{Department of Physics, University at Buffalo, State University of New York, Buffalo, NY 14260, USA}

\author{Farkhad G. Aliev}
\email[e-mail: ]{farkhad.aliev@uam.es}
\affiliation{Departamento F\'isica de la Materia Condensada C-III, Instituto Nicol\'as Cabrera (INC) and  Condensed Matter Physics Institute (IFIMAC), Universidad Aut\'onoma de Madrid, Madrid 28049, Spain}

\begin{abstract}

Spin-orbit coupling (SOC) is a key interaction in spintronics, allowing an electrical control of spin or magnetization and, vice versa, a magnetic control of electrical current. However, recent advances have revealed much broader implications of SOC that is also central to the design of topological states with potential applications from low-energy dissipation and faster magnetization switching to high-tolerance for disorder. SOC and the resulting emergent interfacial spin-orbit fields are simply realized in junctions through structural inversion asymmetry, while the anisotropy in magnetoresistance (MR) allows for their experimental detection. Surprisingly, we demonstrate that an all-epitaxial ferromagnet/MgO/metal junction with a single ferromagnetic region and only a negligible MR anisotropy undergoes a remarkable transformation below the superconducting transition temperature of the metal. The superconducting junction has a three orders of magnitude higher MR anisotropy and could enable novel applications in superconducting spintronics. In contrast to common realizations of MR effects that require a finite applied magnetic field, our system is designed to have two stable zero-field states with mutually orthogonal magnetizations: in-plane and out-of-plane. This bistable magnetic anisotropy allows us to rule out orbital and vortex effects due to an applied magnetic field and identify the SOC origin of the observed MR. Such MR reaches $\sim20\%$ without an applied magnetic field and could be further increased for large magnetic fields that support vortices. Our findings call for revisiting the role of SOC, even when it seems negligible in the normal state, and suggest a new platform for superconducting spintronics.
\end{abstract}

\maketitle

\section{INTRODUCTION}

\begin{figure}[H]
\begin{center}
\includegraphics[width=1\linewidth]{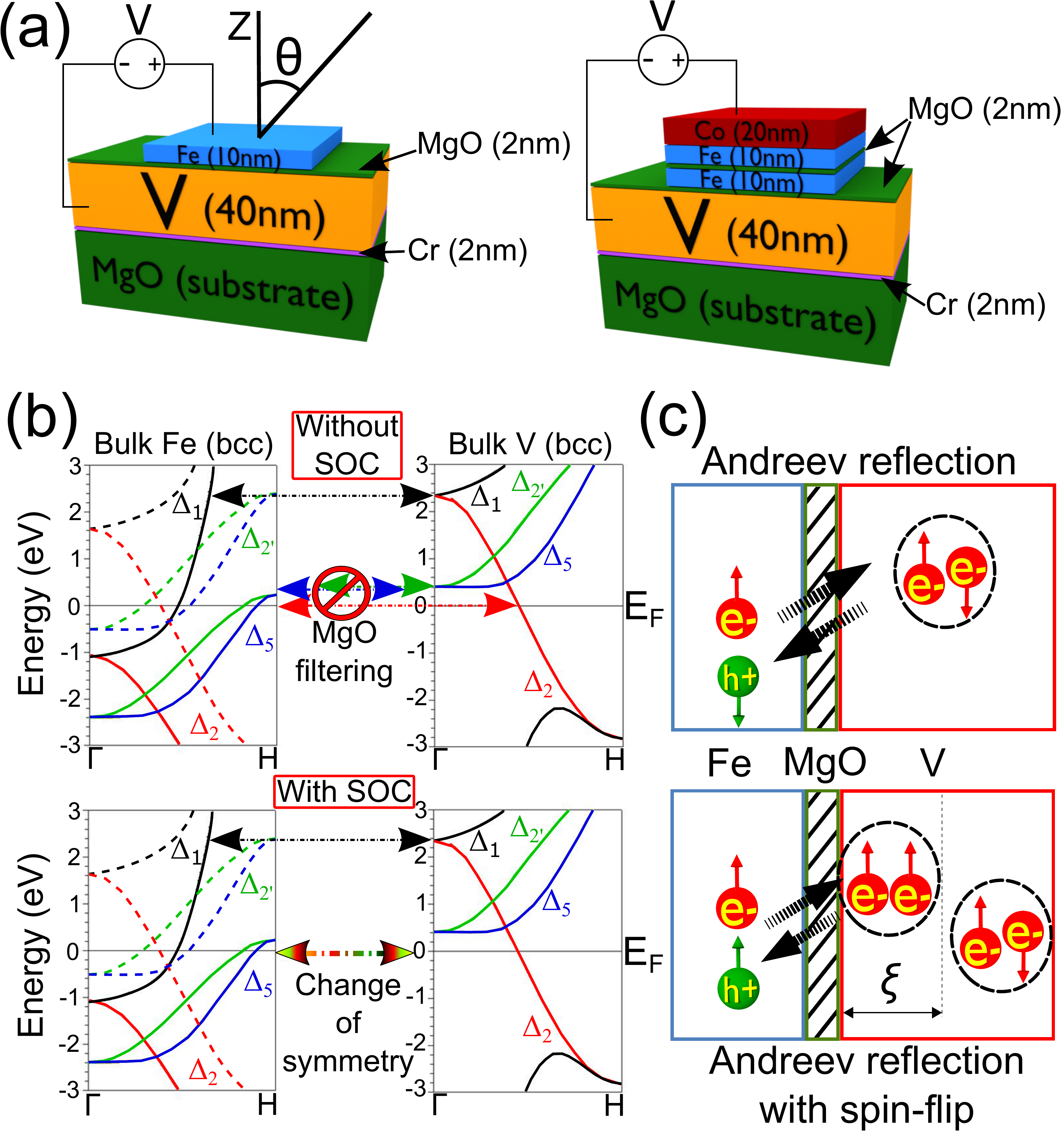}
\caption{Junction geometry, electronic structure, and Andreev reflection. (a) Schematic of Fe/MgO/V and Co/Fe/MgO/Fe/MgO/V junctions, where Co/Fe forms a hard ferromagnet. $\theta$ is the angle between the magnetization, \textbf{M}, and the interface normal. (b) First-principles electronic structure calculations illustrate the orbital symmetry-controlled tunneling across Fe/MgO/V junctions without (with) spin-orbit coupling in the upper (lower) panel, where $\Delta_1,…\Delta_5$  are different orbital symmetries. For Fe(100) majority (minority) bands are given by solid (dashed) lines. (c) Andreev reflection at the Fe/MgO/V interface without (with) spin-flip scattering in the upper (lower) panel. The arrows denote the spin direction of the electrons (e-) and holes (h+), black circles mark the resulting formation of Cooper pairs and $\xi$ the coherence length.}
\label{fig1}
\end{center}
\end{figure}

For over 150 years magnetoresistive effects have provided attractive platforms to study spin-dependent phenomena and enable key spintronic applications~\onlinecite{Zutic2004}. Primarily, spintronics relies on junctions with at least two ferromagnetic layers to provide sufficiently large magnetoresistance (MR). Record room-temperature MR and commercial applications employ magnetic tunnel junctions (MTJs) of common ferromagnets, such as Co and Fe with MgO tunnel barrier~\onlinecite{Parkin2004,Yuasa2004}. Alternatively, MR occurs in single ferromagnetic layers with an interplay of interfacial spin-orbit coupling (SOC). However, in metallic systems this phenomenon, known as the tunneling anisotropic MR (TAMR)~\onlinecite{Fabian2007}, is typically $<1\%$ and precludes practical applications. Here we show experimentally that a negligible MR in an all-epitaxial ferromagnet/MgO/metal junction is drastically enhanced below the superconducting transition temperature of the metal. We find that this peculiar behavior with the role of the interfacial SOC supports the formation of spin-triplet superconductivity which could expand the range of applications in superconducting spintronics~\onlinecite{Singh2015,Linder2015,Eschrig2011} by better integrating superconductivity and ferromagnetism. Building on our experience in fabricating high-quality all-epitaxial MgO-based MTJs~\onlinecite{Guerrero2007,Aliev2007}, we design competing magnetic anisotropies that allow multiple magnetic configuration even at zero applied magnetic field. Furthermore, in our all-epitaxial junctions (see Fig.~\ref{fig1}) the role of SOC is more pronounced as it avoids interfacial wavevector averaging, common to other superconducting junctions.

The quest for spin-triplet superconductivity in ferromagnet/superconductor (F/S) junctions was initially motivated by the long-range proximity effects to overcome the usual competition between superconductivity and ferromagnetism since spin-singlet superconductivity is strongly suppressed by the exchange field~\onlinecite{Buzdin2005,Bergeret2005,Golubov2004}. Such long-range triplet (LRT), with length scales expected only for normal metal (N)/S junctions, supports dissipationless spin currents important for emerging applications in superconducting spintronics~\onlinecite{Singh2015,Linder2015,Eschrig2011}~\onlinecite{Banerjee2014,Baek2014,Gingrich2016,Eschrig2019}. While large MR behavior can occur in superconducting junctions with only spin-singlet superconductivity, the presence of spin-triplet supercurrents could also generate spin torque and alter magnetic configuration or act as a phase battery~\onlinecite{Linder2015,Eschrig2019}.

Unlike the common expectation that large MR implies multiple F regions~\onlinecite{Linder2015}, while LRT requires complex F multilayers, typically relying on noncollinear/spiral magnetization (M)~\onlinecite{Buzdin2005,Bergeret2005,Robinson2010} or half metals~\onlinecite{Singh2015,Keizer2006,Eschrig2003}, we demonstrate that a collinear \textbf{M} in a single F layer provides both large MR and could support LRT since it is accompanied by intrinsic interfacial scattering that mixes spin-singlet and spin-triplet pairing~\onlinecite{Gorkov2001,Bergeret2013,Hogl2015}. A zero-field MR $\sim20\%$ is further enhanced in large applied magnetic fields due to the vortex formation. In an all-epitaxial F/insulator/S (F/I/S) junction, we realize a versatile building block for superconducting spintronics which is compatible with commercial spintronics based on Fe/MgO junctions ~\onlinecite{Parkin2004,Yuasa2004,Cascales2012}. We focus on two types of junctions, shown in Fig.~\ref{fig1}(a). (i) Fe(100)/MgO(100)/V(100) and (ii) F1/I/F2/I/S: Co/Fe/MgO(100)/Fe(100)/MgO(100)/V(100), where vanadium (V) becomes  superconductor below the critical temperature, $T_C=4$ K, while in the latter case magnetically hard F1 (Co/Fe) and soft F2 (Fe) regions provide a versatile control of the \textbf{M}-orientation. These systems illustrate the concept of proximitized materials~\onlinecite{Zutic2019} to design emergent properties, absent in any constituent region of the considered junctions.

\section{EXPERIMENTAL DETAILS}

The normal state transport of these epitaxial junctions, with crystalline MgO and the conserved wave vector parallel to the interfaces, \textbf{k}$_\|=0$, can be understood from the orbital symmetry-controlled tunneling~\onlinecite{Tsymbal2011} across Fe/MgO/V shown in Fig.~\ref{fig1}(b), obtained using full potential calculations in WIEN2k code~\onlinecite{Blaha2001}. Having in view the thickness of the MgO barrier, the main contribution to the tunneling comes from normal incidence at the \textbf{k}$_\|=0$ ($\Gamma$ point). Our ab-initio calculations show that at the Fermi level, $E_F$, the electron states are dominated by different orbital symmetries, $\Delta_1$ in Fe and $\Delta_2$ in V which, in the absence of interfacial scattering that disrupts symmetry, would yield zero low-bias conductance across the MgO(100). Experimentally, in Fe/MgO/V such conductance does not vanish, but is about $10^2$ times smaller than control samples Co/Fe/MgO/Fe and Fe/MgO/Au with similar barrier quality and thickness, as well as similar lateral sizes. This finite low-bias conductance is consistent with the interfacial scattering and the structural inversion asymmetry of a junction~\onlinecite{Zutic2004,Fabian2007}, responsible for the change of symmetry across the MgO by relaxing the symmetry-selection rules. From the absence of symmetry-enforced spin filtering for Fe/MgO/Au junction with see that the suppression of conductance in Fe/MgO/V is dominated by the spin filtering, rather than by the value of the barrier strength, which is largely unchanged.  While fabricated Fe/MgO are not strictly half-metallic, through $\Delta_1$ symmetry they provide a very large tunneling spin polarization ($85\%$ in Ref.~\onlinecite{Parkin2004}), which is also the origin of the huge TMR in Fe/MgO-based MTJs~\onlinecite{Parkin2004,Yuasa2004}.

The MTJ multilayer stacks have been grown by molecular beam epitaxy (MBE) on (100) MgO single crystal substrates in a chamber with a base pressure of $5\times10^{-11}$ mbar following the procedure described in Ref.~\onlinecite{Tiusan2007}. First, a 10 nm thick seed MgO anti-diffusion underlayer was grown on the substrate to trap the residual C. Then, the V electrode is deposited at room temperature and further annealed for flattening at $T>500 C^0$. The choice of the temperature and annealing time was monitored by RHEED pattern analysis. Then MgO insulating barrier layer was epitaxially grown by e-beam evaporation with 2 nm thickness, precisely controlled at sub-monolayer scale, by in-situ RHEED intensity analysis. This MgO thickness is in the asymptotic transport regime~\onlinecite{Butler2001}, where the orbital symmetry filtering across the barrier leads to a main contribution to the \textbf{k}$_\|=0$ tunneling. For a given thickness, MgO barrier is typically not as strong as for aluminum oxide. In our epitaxially-grown MgO there is an additional reduction of the effective barrier to $\sim0.3-0.4$ eV for a low-bias transport, consistent with the transport measurements on epitaxial Fe/MgO-based MTJs in Ref.~\onlinecite{Yuasa2004}. In contrast, the barrier is $\sim1$ eV higher for polycrystalline non-epitaxial Fe/MgO-based MTJs, studied in Ref.~\onlinecite{Parkin2004}. 
   
The epitaxial growth sequence is continued for all the other layers, leading to the single crystal MTJ stacks in which crystalline symmetry across the stack ensures the symmetry and \textbf{k}$_\|$ conservation of the Bloch electron function, allowing a very good agreement between the theory and experiment. For F1/I/F2/I/S junctions [Fig.~\ref{fig1}(a), right], the hard ferromagnet F1 is formed from a 10 nm thick Fe layer, epitaxially grown on the top of the MgO, and a 20 nm Co layer on top of it. After the MBE growth, all the MTJ multilayer stacks are patterned in micrometer-sized square junctions by UV lithography and Ar ion etching, controlled step-by-step in-situ by Auger spectroscopy. Details of conductance measurements and the vector magnetic field control are in Refs.~\onlinecite{Cascales2012,Martinez2018}. A unique feature of our devices is the control of the remanent \textbf{M}-direction of the magnetically soft electrode interfacing S. Such a control is possible due to presence of competing perpendicular and in-plane magnetic anisotropies in 10 nm thick epitaxial Fe layers interfacing MgO~\onlinecite{Martinez2018}. 

\section{RESULTS}

The transport in the superconducting state is distinguished by the Andreev reflection, providing the microscopic mechanism for proximity-induced superconductivity~\onlinecite{Zutic2004,Buzdin2005,Bergeret2005}. During conventional Andreev reflection shown in Fig.~\ref{fig1}(c) (upper panel), an electron is reflected backwards and converted into a hole with opposite charge and spin. This implies the doubling of the normal state conductance~\onlinecite{Blonder1982} since two electrons are transferred across the interface into the S region where they form a spin-singlet Cooper pair. With unequal contributions of minority and majority spins in the F region, not all electrons can find the partner of opposite spin to undergo the Andreev reflection. Therefore, the spin polarization of the F region can be studied through the suppression of Andreev reflection and the resulting low-bias conductance. While this procedure usually relies on point contacts~\onlinecite{Soulen1998}, it was also applied for F/I/S tunnel junctions~\onlinecite{Parker2002} with additional suppression of Andreev reflection due to potential scattering at an insulator. In both cases, it is important to decouple the role of conductance suppression due to spin polarization from that arising from interfacial scattering~\onlinecite{Nadgorny2012,Ren2007,Miyoshi2005,Sangiao2011,Yates2013}. In contrast to this conventional Andreev reflection, a spin-active interface with interfacial spin-flip scattering also can yields Andreev reflection with an equal spin of electrons and holes~\onlinecite{Zutic1999}, responsible for a spin-triplet Cooper pair shown in Fig.~\ref{fig1}(c) (lower panel). Since Andreev reflection is the origin of superconducting proximity effects~\onlinecite{Zutic2004,Buzdin2005,Bergeret2005,Hogl2015}, one expects that the resulting interfacial spin-triplet correlations decay away from the interface both inside the S and F regions. The decay length inside the F region is longest for spin-triplet (equal spin) correlations~\onlinecite{Linder2015,Eschrig2011,Buzdin2005,Bergeret2005}.

\begin{figure}[H]
\begin{center}
\includegraphics[width=1\linewidth]{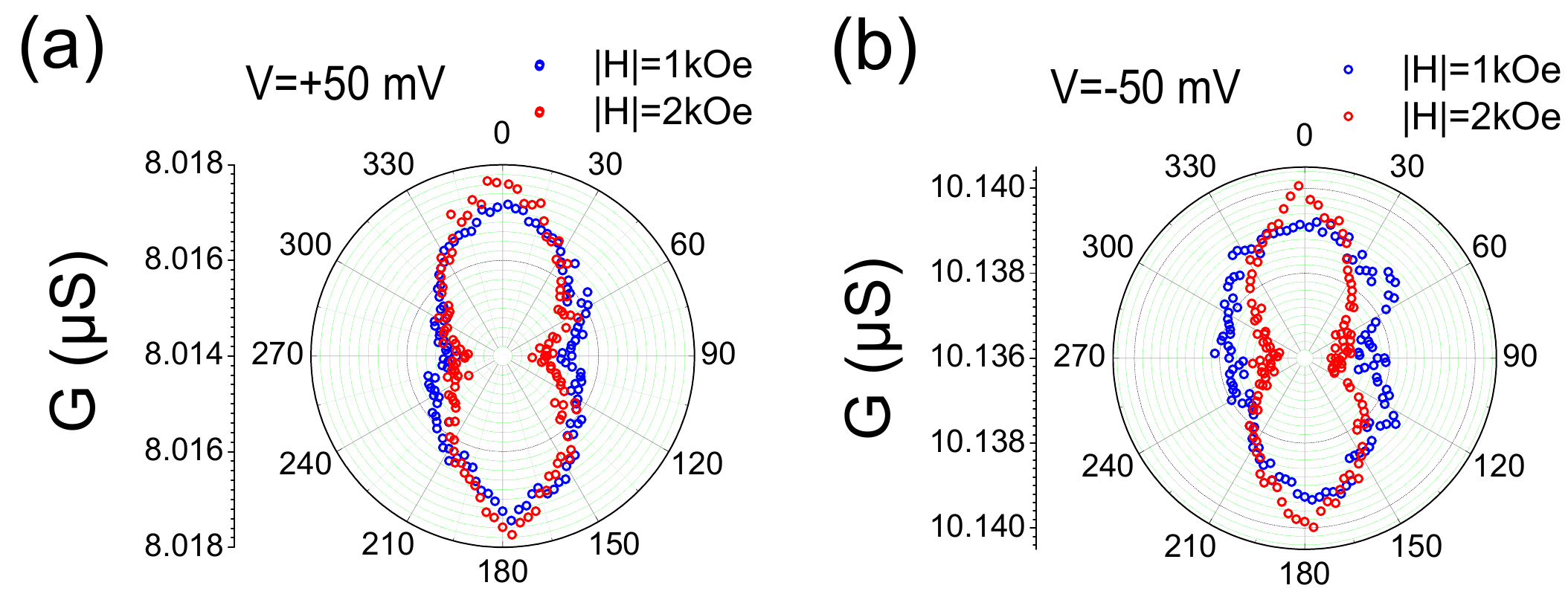}
\caption{Angular dependence of conductance for an out-of-plane rotation of magnetization, \textbf{M}. The conductance is measured for \textbf{M} at a polar angle $\theta$ normal to the interface at $T=0.3$ K and applied bias $V=+50$ mV (a) and $V=-50$ mV (b), and applied fields $H=1$ kOe and 2 kOe. Since the voltage greatly exceeds $\Delta$ for vanadium, the resulting magnetic anisotropy can be used to extract the magnitude of normal state TAMR $\sim0.01\%$ and $0.02\%$ for $H=1$ kOe and 2 kOe, respectively.}
\label{fig2}
\end{center}
\end{figure}

While the bias-dependent conductance, $G(V)$, can indicate the presence of interfacial SOC, similar $G(V)$ also arise from the \textbf{k}-independent interfacial spin-flip scattering~\onlinecite{Hogl2015,Zutic1999} due to local exchange coupling. Instead, a unique fingerprint of interfacial SOC is the magnetic anisotropy of G(V). Within the normal state, $T>T_C$ or, equivalently, for bias above the superconducting gap $V\gg\Delta$, the TAMR in F/I/N junctions can reveal such SOC through \textbf{M}-dependent G~\onlinecite{Fabian2007}. This can be seen in Fig.~\ref{fig2}, where the angle-dependent conductance, $G(\theta)$, expressed using $\theta$ from Fig.~\ref{fig1}(a). However, the magnitude of the observed $\text{TAMR}=[G(0)-G(\pi/2)]/G(\pi/2)\sim0.01\%$, with $G(\theta)$, while reproducible, is too small for any practical use.

\begin{figure}[H]
\begin{center}
\includegraphics[width=1\linewidth]{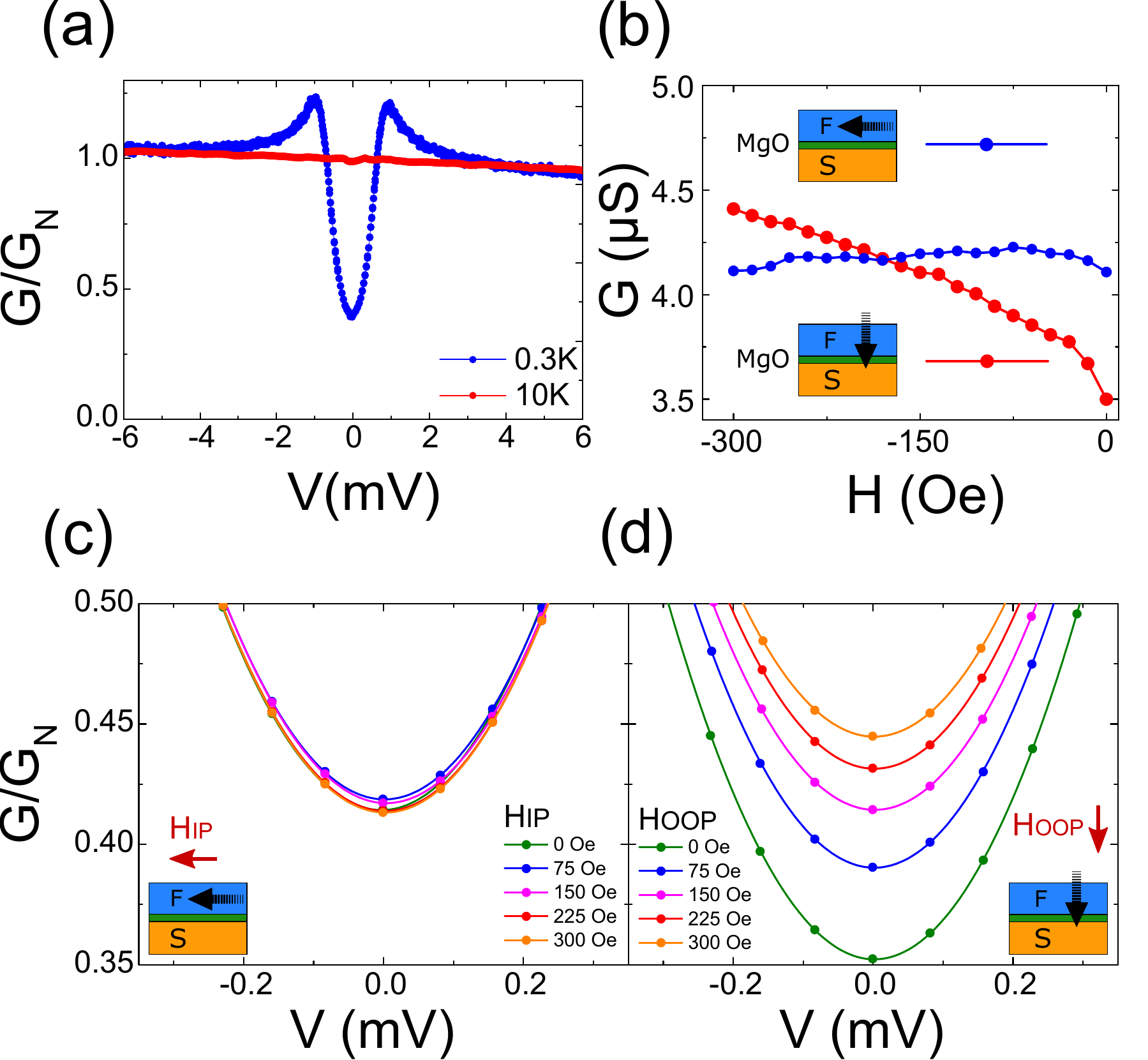}
\caption{Conductance dependence with bias and applied magnetic field, $H$. (a) Typical low-field conductance in Fe/MgO/V junctions at $H=0$ Oe, below and above $T_C$. (b) Evolution of the zero bias conductance with in-plane and out-of-plane $H$ at $T = 0.3$ K. The two remanent perpendicularly oriented magnetic states (black arrows) with different $G(0)$ reveal $\sim17\%$ conductance anisotropy (MAAR) at $H=0$. (c), (d) The low-bias dependence of the conductance at $T = 0.3$ K shows a different behavior with an in-plane and an out-of-plane $H$. The conductance is normalized to its normal-state value, $G_N=G(V=3 \text{ mV})$.}
\label{fig3}
\end{center}
\end{figure}

Motivated by the prediction that MR can be enhanced in the superconducting state~\onlinecite{Hogl2015}, we explore in the same junction the resulting F/I/S conductance anisotropy. The corresponding out-of-plane magnetoanisotropic Andreev reflection (MAAR)~\onlinecite{Hogl2015}, is the superconducting analog of the TAMR, defined as
\begin{equation}
\text{MAAR}=\dfrac{G(0)-G(\theta)}{G(\theta)},
\end{equation}
where $\theta$ is the \textbf{M} angle with the interface normal (see Appendix A) which compares TAMR and MAAR). A useful reference information about the transport across the junction is provided in Fig.~\ref{fig3}(a) by comparing a typical zero-field $G(V)$ below and above $T_C$, for $T=0.3$ K ($T/T_C=0.08$) and $T=10$ K, respectively. Just as in a widely-used one-dimensional BTK model for $G(V)$ of N/I/S junctions, excluding SOC and the spin polarization~\onlinecite{Blonder1982}, a common description generalizing it to F/I/S junctions~\onlinecite{Hogl2015,Soulen1998,Parker2002,Nadgorny2012,Ren2007} relies on a dimensionless interfacial barrier parameter $Z$ (see Appendix). The normal-state transparency for the BTK model, $1/(1+Z^2)$, establishes a simple parameterization for the perfectly transparent junction with $Z=0$ and the tunnel junction with $Z\gg1$ with vanishing transparency. The lowering of conductance is thus directly connected to the interfacial barrier~\onlinecite{Blonder1982}. The normalized conductance is consistent with an intermediate junction transparency (see Appendix). It is lower than for point contacts where the Andreev reflection is used to probe the spin polarization of ferromagnets~\onlinecite{Soulen1998,Nadgorny2012}, but higher than for typical spin-polarized quasiparticle tunneling, pioneered in F/I/S junctions by Tedrow and Meservey~\onlinecite{Zutic2004}. In the latter case~\onlinecite{Parker2002}, seen also for polycrystalline, non-epitaxial, Fe/MgO/Al junctions~\onlinecite{Parkin2004}, the tunneling conductance is more suppressed at lower bias and displays a larger peak near $V\sim\Delta$ expected from the BCS density of states.

While the absolute conductance could be inferred from the BTK model~\onlinecite{Blonder1982}, such that in the absence of interfacial barrier $Z=0$ case recovers the value of the Sharvin conductance~\onlinecite{Zutic2004}, various deviations have been known for a while~\onlinecite{Ren2007}. In F/I/S junctions the conductance can be several orders of magnitude lower than predicted by the BTK model. However, the relevance of BTK-like description is still ascertained by recognizing that it accurately describes the normalized conductance spectra~\onlinecite{Ren2007}. In our junctions reduced absolute conductance, as compared to the BTK predictions, can be inferred from Fig.~\ref{fig3}(b). This is consistent with the epitaxial character of our F/I/S junctions. Even if the interfacial barrier is removed, symmetry-enforced spin filtering would not a give a simple limit of Sharvin conductance which neglects the relevant spin-related electronic structure and the role of SOC on spin filtering. In our control all-epitaxial N/I/S sample Au(15 nm)/ MgO(2 nm)/V(40 nm), grown with the same method as the samples from Fig.~\ref{fig1}, replacing Fe by Au leads to the 1000-fold larger conductance. This striking increase cannot be explained by $\sim0.2$ eV reduction in the interfacial barrier, but rather shows the importance of the symmetry-imposed spin filtering absent with Au.

To minimize $H$-dependent MAAR effects, potentially leading to anisotropy because of the vortex distribution in the superconductor, in Fig.~\ref{fig3}(b) we identify the lower bound of MAAR by performing remanent measurements at $H=0$. While such a goal to realize multiple \textbf{M}-orientation, stable at $H=0$, is impossible in typical F/I/S junctions, a unique feature in the design of our structures are their multiple non-volatile states, depicted in Fig.~\ref{fig3}(b), resulting from the competing perpendicular and in-plane anisotropy~\onlinecite{Martinez2018}. We first applied and removed in-plane saturation field of $\sim4$ kOe in the required direction. In the second stage, the field dependences of conductance have been measured by departing from those different remanent states. The corresponding measurements of $G(0)$ and $G(\pi/2)$ at V$=0$ confirms a giant increase in the magnetic anisotropy of the superconducting state with the MAAR of$\sim17\%$. 

An applied magnetic field, $H$, leads to the vortex formation and a gradual suppression of superconductivity in V, destroyed at a critical field $H=H_{C2}$. For our thin V film, the measured $H_{C2}$ is anisotropic: $H_{C2,\perp}=3.5$ kOe and $H_{C2,\|}=12$ kOe. From a given range of $H$ in Figs.~\ref{fig3}(c) and (d) we can also infer such$H_{C2}$ anisotropy from low-bias $G(V)$. There is only a negligible $G(V)$ change for an in-plane $H$, since $H/H_{C2,\|}\ll1$. For an out-of-plane $H$ we see a larger change, an increase of $G(V)$ is expected for a suppression of superconductivity. 

From $H=0$ data in Figs.~\ref{fig3}(b)-(d) we can rule out that MAAR is a trivial effect due to the induced vortices with an applied $H$. The observed MAAR also substantially exceeds the possible effect of the fringing fields from the Fe region on V film separated by the MgO, as evaluated from micromagnetic simulations using MuMax$^3$ code~\onlinecite{Vansteenkiste2014} ($<20$ Oe in the central region and $<150$ Oe near the edges, still less than $H_{C1}$ for V~\onlinecite{Alekseevskii1976,Sekula1972}). We have verified on a control all-epitaxial Au/MgO/V junction than by applying perpendicular $H$-field in the amount of these calculated fringing fields changes the measured conductance by less than $1\%$. If vortices at $H=0$ were formed by such fringing fields, their effect on V film would be stronger for an out-of-plane \textbf{M} and thus show a stronger suppression of superconductivity as reflected by a larger $G(V=0)$ which would be closer to the normal state value, $G_N$. Instead, we see from Figs.~\ref{fig3}(c) and (d) an opposite effect: $G(V=0)$ is smaller for an out-of-plane \textbf{M}, inconsistent with a stronger suppression of superconductivity and MAAR dominated by vortices. 

Remarkably, compared to Fig.~\ref{fig2}, in Fig.~\ref{fig3} there is a huge increase in the observed anisotropy, seemingly inconsistent with only a very small interfacial SOC responsible for TAMR. To reconcile such an increase in magnetic anisotropy it is useful to recognize the sensitivity of Andreev reflection to interfacial SOC for a highly spin-polarized ferromagnet~\onlinecite{Zutic2004}. In the limiting case of a complete spin polarization with the absence of minority spins no conventional Andreev reflection [Fig.~\ref{fig1}(c)] is expected and thus $G(V<\Delta)=0$. However, with interfacial SOC, a spin-flip [equal spin, see Fig.~\ref{fig1}(c)] Andreev reflection is possible, supporting the triplet superconductivity.

\section{MODEL AND DISCUSSION}

\begin{figure}[H]
\begin{center}
\includegraphics[width=1\linewidth]{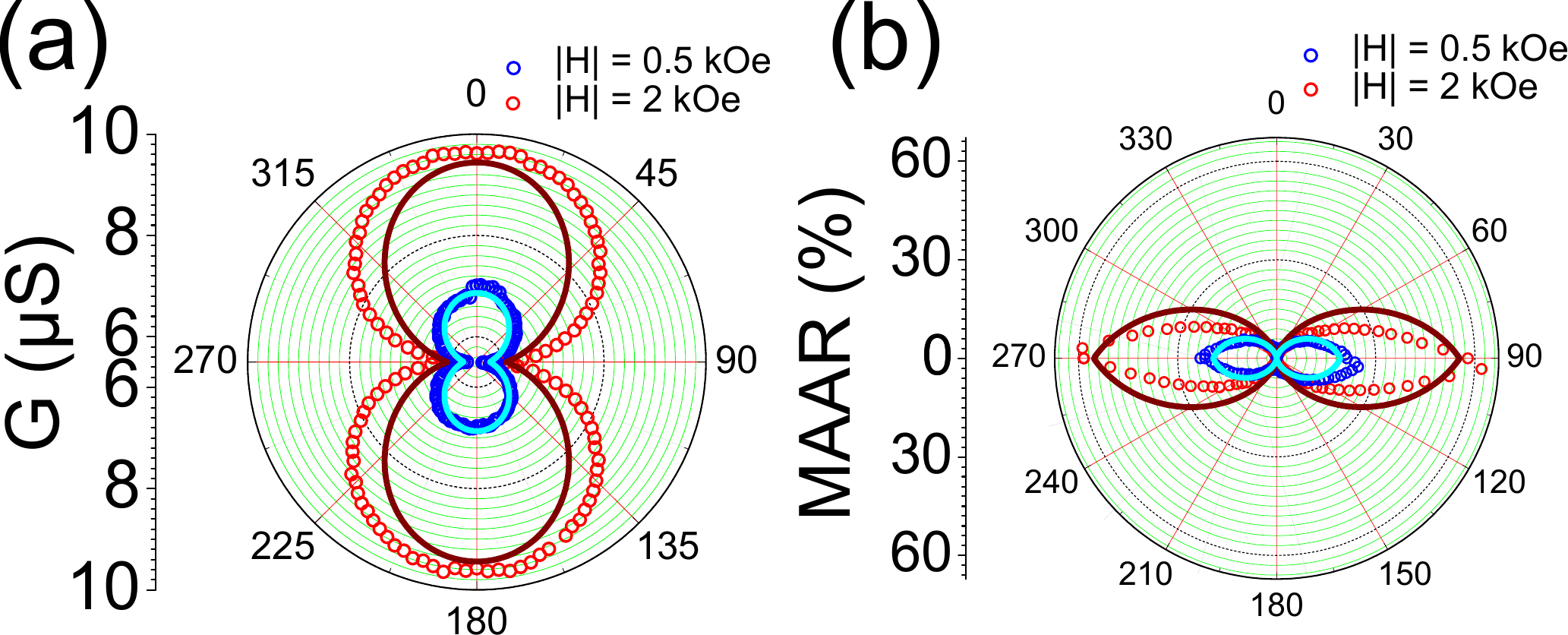}
\caption{Magnetoanisotropic Andreev reflection (MAAR). (a) Out-of-plane zero bias conductance anisotropy at $T=0.3$ K, $H=0.5$ kOe (blue dots) and $H=2$ kOe (red dots), compared to our phenomenological model including magnetic field effects (solid lines) with fitting parameters (see text) $G_0=6.01$ $\mu$S, $A=-0.0415$ $\mu$S, and $B=-0.224$ $\mu$S. (b) The same approach for out-of-plane MAAR.}
\label{fig4}
\end{center}
\end{figure}

Orbital symmetry-controlled tunneling turns Fe/MgO into a source of highly spin-polarized carriers~\onlinecite{Zutic2004,Parkin2004,Yuasa2004}, while the structural inversion asymmetry in Fe/MgO/V junction leads to an interfacial Rashba SOC~\onlinecite{Zutic2004,Fabian2007} with the field $w_R=(\alpha k_y,-\alpha k_x)$, where $\alpha$ is the interfacial Rashba SOC parameter and $\text{\textbf{k}}_\|=(k_x,k_y)$ the in-plane wave vector. While the presence of interfacial SOC can be already inferred from $G(\theta)$ in Fig.~\ref{fig4}(a), it is also helpful to examine the corresponding out-of-plane MAAR~\onlinecite{Hogl2015} from Eq. (1), shown in Fig.~\ref{fig4}(b). A further support for the interfacial SOC comes from $G(0)$ in Fig.~\ref{fig3}(a): its suppression expected for a highly-spin polarized ($\sim70\%$) Fe/MgO interface~\onlinecite{Zutic2004} is rather small, as compared with $G(V\gg\Delta)$ pointing to unconventional Andreev reflection as the main contribution to $G(0)$. For $V\gg\Delta$\:shot noise measurements (see Appendix B) confirm electron tunneling, excluding the presence of pinholes as the origin of a relatively high $G(0)$.    

Although the measured MAAR and its relation to the interfacial SOC appear to be in qualitative agreement with previous theoretical predictions~\onlinecite{Hogl2015}, sizable magnetic field effects result in a large discrepancy with respect to the magnitude and angular-dependence of the MAAR(see appendix). In F/I/S junctions the rotation of \textbf{M}, in a finite $H$, generally yields two additional effects which were previously not considered~\onlinecite{Hogl2015}: Orbital contributions and $H$-dependent suppression of superconductivity. The orbital effects are related to the Lorentz force and cyclotron orbits which we include phenomenologically~\onlinecite{Fabian2007} by shifting the initial \textbf{k}$_\|$ in the reflection probability and the SOC field, resulting in the conductance (see Appendix),
\begin{equation}
G^j(V,\theta,H)={G_0}^j + {G_\alpha}^j(1 - \cos{2\theta})
\end{equation}
separated into SOC independent (dependent) ${G_0}^j ({G_\alpha}^j)$ part, with ${G_\alpha}^j={g_1}^j\alpha H+ {g_2}^j\alpha^2$, where $g_{1,2}^j$ are real and SOC independent coefficients, while index $j=N,S$ labels the normal, superconducting contributions. The other effect of an applied $H$, as discussed above, is the vortex formation and a gradual suppression of superconductivity, destroyed at $H=H_{C2}(\theta)$~\onlinecite{Miyoshi2005}. From our measured values $H_{C2,\perp}$ and $H_{C2,\|}$, the angular dependence of $H_C$ in a V thin film can be described by Tinkham’s formula, $(H_{C2}(\theta)/H_{C2,\|})^2\sin^2{\theta}+(H_{C2}(\theta)/H_{C2,\perp})|cos{\theta}|=1$ (see Appendix and  Ref.~\onlinecite{Tinkham1963}). We define a dimensionless field $h(\theta)=H/H_{C2}(\theta)$ that quantifies the creation of vortices introducing a normal contribution to conductance, as well as the suppression of its superconducting part (see Ref.~\onlinecite{Miyoshi2005} and Appendix),           
\begin{equation}
G(V,\theta,H)=h(\theta){G}^N + (1 - h(\theta))G^S
\end{equation}
In our analysis of the $G(V=0)$ data for a complete rotation of \textbf{M} from Fig.~\ref{fig4}(a), we apply the phenomenological model described by Eqs. (2) and (3), where we assume $G^N=2G^S$, as in the measured $G(V=0)$ from Fig.~\ref{fig10}(b) in Appendix  and omit index S. \:From $G(\theta)$ data at $H=0.5$ kOe and 2 kOe, we first determine $G_0$ at $\theta=0$ and then solve for the two parameters $A=g_1\alpha H_{C,\perp}$and $B=g_2\alpha^2$ at $\theta=3\pi/2$.\:While we use the measured data of only these two angles, we recover a good agreement in both $G$ and MAAR over the full range of $\theta$\: shown in Figs.~\ref{fig4}(a) and (b) with solid lines. This also yields an expected $G_0\gg G_\alpha$ from our model, corroborating the role of interfacial SOC on both the giant MAAR of $\sim20\%$ ($\sim60\%$) at 0.5 kOe (2kOe)~\onlinecite{Zutic1999}. Our agreement between our phenomenological model and the data for $G(\theta)$ and $\text{MAAR}(\theta)$ is even better for a smaller $H=0.5$ kOe, since this model contains only the leading corrections to the applied $H$.  

\begin{figure}[H]
\begin{center}
\includegraphics[width=1\linewidth]{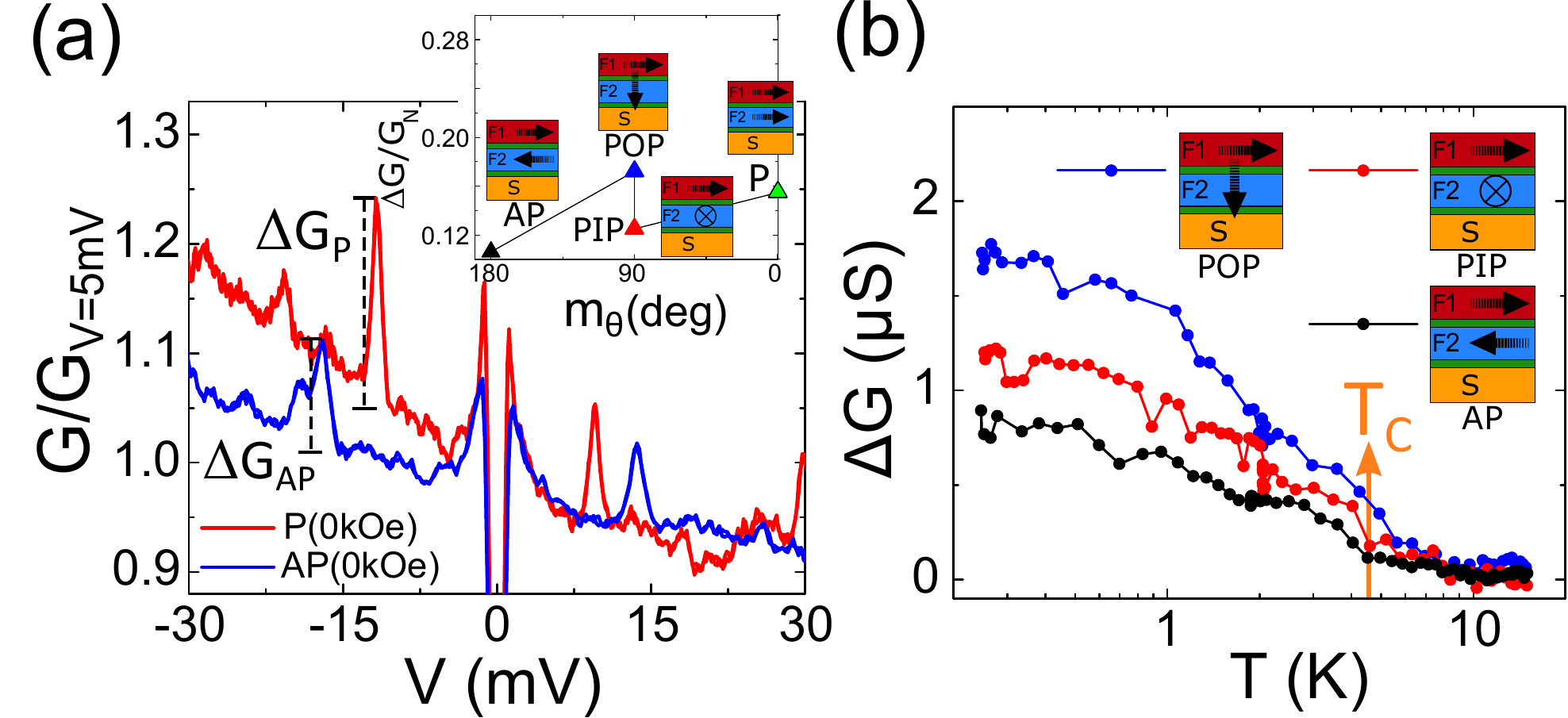}
\caption{Conductance characterization of F1/I/F2/I/S junctions at remanence.  (a) Bias dependence of the conductance normalized by its above the gap value at $V=5$ mV and $T=0.3$ K for parallel (P) and antiparallel (AP) \textbf{M}-orientation (black arrows), $\Delta G$ denotes above the gap largest conductance anomalies. Inset: the variation of the normalized largest $\Delta G$ (observed around -15 mV) with the \textbf{M}-orientation, P, AP, perpendicular out-of-plane (POP), and perpendicular in plane (PIP). Top F region is magnetically hard (Fe/Co) and the bottom (Fe) is soft. (b) The temperature dependence of the amplitude of the strongest conductance anomaly for different magnetic states POP, PIP, and AP. }
\label{fig5}
\end{center}
\end{figure}

The nonvolatile control of \textbf{M} and modifications of superconductivity can be extended in F1/I/F2/I/S junctions, as shown in Fig.~\ref{fig5}, to realize four different states: parallel (P), antiparallel (AP), perpendicular in plane (PIP), and perpendicular out-of-plane (POP). While similar spin-valve structures are common to superconducting spintronics~\onlinecite{Singh2015,Linder2015,Eschrig2011}, within our platform they reveal further opportunities due to the presence of additional nonvolatile configurations that are realized at $H=0$. Conductance variations above the gap in Fig.~\ref{fig5} resemble findings in F/S-based structures from Ref.~\onlinecite{Villegas2012}, explained by LRT, and show peculiar trends that could motivate further theoretical studies of interfacial SOC in all-epitaxial junctions. These observed conductance anomalies are periodic (see Appendix showing typical CAs in a wider bias range), disappear slightly above $T_C$ and are sensitive to the relative \textbf{M}-orientation. They are likely to arise from the quasiparticle interference inside a 10 nm thick Fe surrounded by two MgO layers, while the presence of the two periods above the gap was explained in Ref.~\onlinecite{deJong1995} due to shorter and longer coherent electron-hole trajectories. Different magnitudes in conductance variations for two perpendicular configurations, with \textbf{M} in F2 being in-plane vs out-of-plane, suggest unconventional pairing. Unlike the expected weakest suppression of spin-singlet superconductivity with antiparallel \textbf{M}~\onlinecite{Buzdin2005}, the amplitude of conductance anomalies in the antiparallel configuration is suppressed for all $ T<T_C$, as compared to the other two \textbf{M}-orientations. These variations remain practically unchanged if normalizing at -5 mV and enhance if the absolute values are compared. Since the changes in the normalized conductance below $T_C$ can reach up to $50\%$ (between AP and POP, Fig.~\ref{fig5} insert), multiple F regions can support previously unexplored MR effects. Another extension of our all-epitaxial junctions would be to include the second superconducting layer and explore the modification of Josephson effect from the interplay between the interfacial SOC and different magnetic configurations. 

\section{CONCLUSIONS}

To the best of our knowledge, all-epitaxial F/I/S junctions have not been previously experimentally realized. This platform could both stimulate theoretical studies beyond the BTK-like description and include the symmetry-controlled spin filtering as well as enable experimental progress towards the applications in superconducting spintronics. A common requirement to realize SOC-driven emergent phenomena, from topological states to LRT is typically SOC that is already inherently strong in the normal state, for example, implemented with heavy elements and in narrow-band semiconductors~\onlinecite{Eschrig2019,Jeon2018,Banerjee2018}. In contrast, the platform we have studied reveals a peculiar superconducting behavior even when a rather weak SOC in the normal state leads to a negligible magnetic anisotropy. Future experiments in this platform could also test a recent prediction about the $\pi/4$ change of the in-plane magnetic easy axis below $T_C$~\onlinecite{Johnes2019} as a further support for the efficient generation of the spin triplet Cooper pairs, excluding alternative scenarios were our observed conductance changes with \textbf{M}-direction would be attributed to trapped vortices or manifestations of a quasiparticle transport. We expect that revisiting Fe/MgO-based junctions, widely used in commercial spintronic applications, will provide an opportunity to use multiple proximity effects: spin-orbit, magnetic, and superconducting, to transform a large class of materials and realize unexplored phenomena~\onlinecite{Zutic2019}.

\section{ACKNOWLEDGEMENTS}

The work at Madrid was supported in part by Spanish MINECO (MAT2015-66000-P, RTI2018-095303-B-C55, EUIN2017-87474, MDM-2014-0377) and Comunidad de Madrid (NANOMAGCOST-CM P2018/NMT-4321). C.T. acknowledges ''EMERSPIN´´ grant ID PN-III-P4-ID-PCE-2016-0143, No. UEFISCDI:22/12.07.2017. J.P.C. acknowledges support from the Fundaci\'on S\'eneca (Regi\'on de Murcia) postdoctoral fellowship (19791/PD/15). The work at Regensburg was supported by the DFG SFB 689, International Doctorate Program Topological Insulators of the Elite Network of Bavaria, and GRK 1570. The work in Nancy was supported by CPER MatDS and the French PIA project 'Lorraine Universit\'e d'Excellence', reference ANR-15-IDEX-04-LUE. Experiments were performed using equipment from the TUBE. DAUM was funded by FEDER (EU), ANR, R\'egion Grand Est and Metropole Grand Nancy. The work at Detroit was supported by the US ONR N000141712793 and DARPA DP18AP900007. The work at Buffalo was supported by the Department of Energy, Basic Energy Sciences (Grant No. DE-SC0004890).

\section*{APPENDIX A: COMPARISON BETWEEN TAMR AND MAAR}

Tunneling anisotropic magnetoresistance (TAMR) and magnetoanisotropic Andreev reflection (MAAR) provide important information about the anisotropy of the transport properties  in the normal and superconducting state, respectively (see the main text). They have the same functional form that can be expressed in terms of the angle-dependent conductance in the normal and superconducting state,~\onlinecite{Hogl2015}
\begin{equation}
\text{TAMR (MAAR)}=\dfrac{G(0)-G(\theta)}{G(\theta)},
\end{equation}
where the angle $\theta$ is measured between the magnetization, {\bf M}, and the interface normal of the  junction, such that an out-of-plane rotation of {\bf M} is considered. 

\begin{figure}[H]
\begin{center}
\includegraphics[width=1\linewidth]{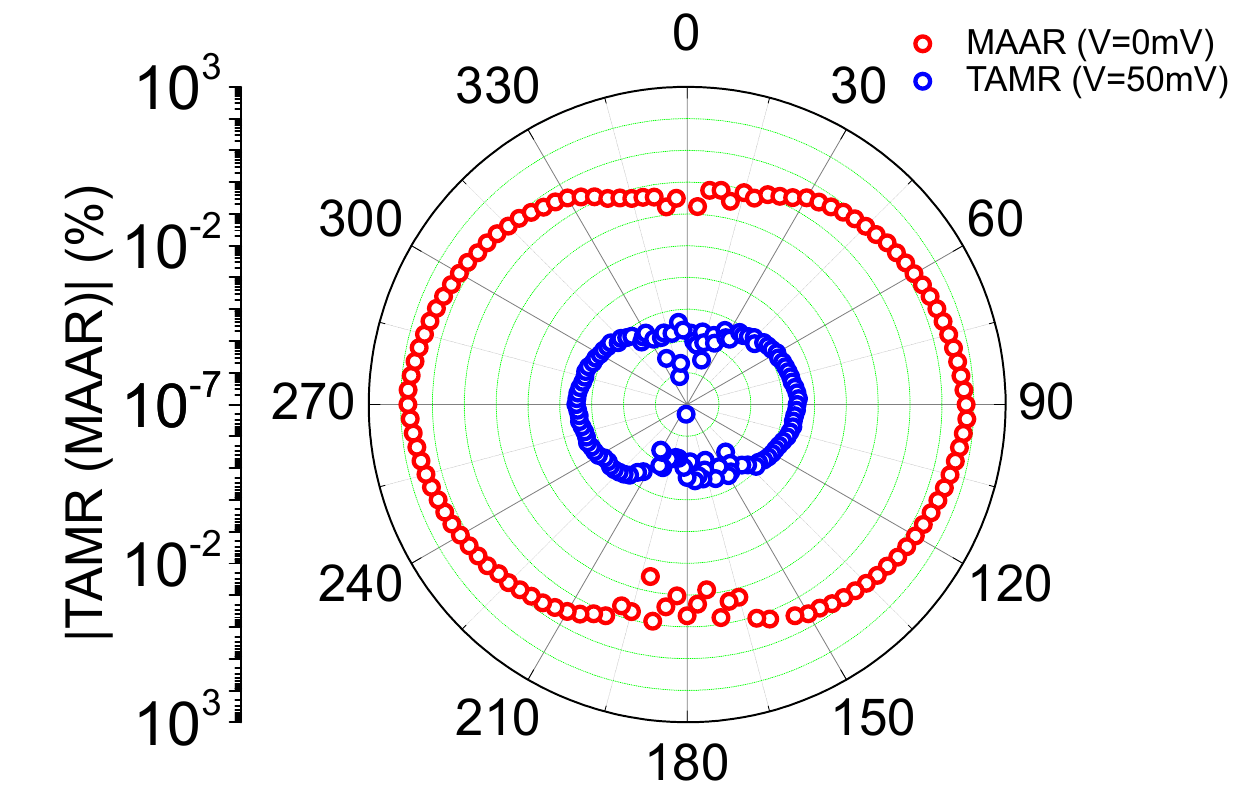}
\caption{A comparison between TAMR and MAAR. Measurements were performed at temperature $T=0.3$ K and an applied bias of $V=50$ mV (TAMR) and 0 mV (MAAR) with an applied magnetic field $H=2$ kOe.  The vanishing TAMR and MAAR for the angle $\theta=0^\circ$ are excluded to avoid a diverging logarithm.}
\label{fig6}
\end{center}
\end{figure}

To provide a direct comparison of the TAMR and MAAR, which for the same sample have several orders of magnitude different amplitudes, we show their logarithmic plots and exclude the $\theta=0^\circ$ results (to avoid a diverging logarithm). The corresponding results are shown in Fig.~\ref{fig6} for an all-epitaxial ferromagnet/insulator/superconductor (F/I/S) Fe/MgO/V junction, depicted in Fig.~\ref{fig1} of the main text. The distinction between the normal and superconducting state, and thus between the TAMR and MAAR, is realized by changing the applied bias, $V$: above the superconducting gap of vanadium $\lesssim 1$ meV ($V\gg \Delta$) for TAMR and $V=0$ for MAAR, while the sample temperature $T=0.3$ K and applied magnetic field $H=2$ kOe are kept fixed.

\section*{APPENDIX B: SHOT NOISE CHARACTERIZATION OF BARRIER QUALITY}

\begin{figure}[tbp]
\begin{center}
\includegraphics[width=0.8\linewidth]{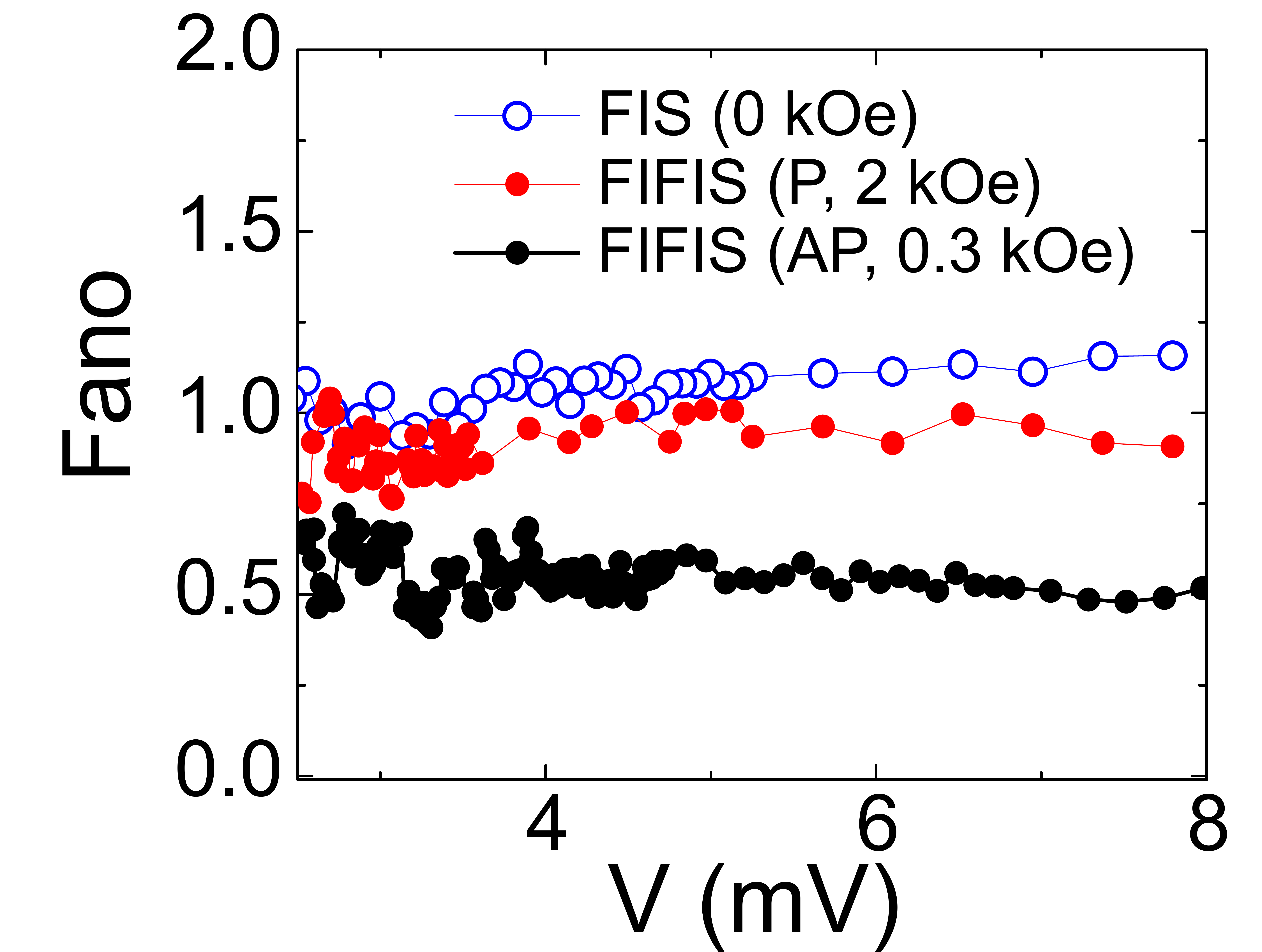}
\caption{Bias dependence of the Fano factor for F/I/S and F/I/F/I/S junction in the P state ($H=2$ kOe) and AP state ($H=0.3$ kOe) measured at 
$T=0.3$ K 
for bias exceeding the superconducting gap ($\Delta \lesssim 1$ meV).}
\label{fig7}
\end{center}
\end{figure}

The shot noise is an intrinsic quantum property arising from the discreteness of the charge carriers.~\onlinecite{Blanter2000} It is convenient to express its dimensionless form as a Fano factor,  noise-to-current ratio, since it can attain universal values that are independent of the details of the system.~\onlinecite{Tsymbal2011,Cascales2012} Such a Fano factor is used to characterize magnetic tunnel junctions (MTJs) and the corresponding tunneling magnetoresistance (TMR) since it depends on the relative orientation of the magnetization in different ferromagnets.~\onlinecite{Tsymbal2011} For example, for MTJs with a single barrier region the Fano factor approaches one in the tunneling limit.~\onlinecite{Tsymbal2011,Cascales2012} The observation of such a value in Fig.~\ref{fig7} indicates direct tunneling through a pinhole free barrier.  
With two possible magnetic states in double barrier F/I/F/I/S junctions, the expected shot noise is evaluated using a model of sequential tunneling~\onlinecite{Cascales2012,Szczepanski2013} The influence of resonant tunneling is not included. The measured Fano factor for the parallel (P) 
and antiparallel AP magnetic configurations could be used to calculate the TMR ratio. The corresponding value shows a good agreement with the measured TMR (in the limit of strong spin relaxation). 

Following the calculation of shot noise in the presence of spin relaxation,~\onlinecite{Szczepanski2013} the Fano factor is given by
\begin{equation}
F=\dfrac{R_{2\uparrow}R_{2\downarrow}(R_{1\uparrow}+R_{1\downarrow})^2+R_{1\uparrow}R_{1\downarrow}(R_{2\uparrow}
+R_{2\downarrow})^2}{[R_{1\uparrow}R_{1\downarrow}(R_{2\uparrow}+R_{2\downarrow})+R_{2\uparrow}R_{2\downarrow}(R_{1\uparrow}+R_{1\downarrow})]^2},
\end{equation}                                                                                                 
where \textit{R} is the partial resistance of each of the two barriers (indices 1,2) and of each of the spin directions (up $\uparrow$ and down $\downarrow$).

For F/I/F/I/S junction the two barriers separate different systems. The normal state resistance of F/I/S system (for $V>\Delta$), $R_{FIS}$ is largely independent of the relative magnetic configuration of the two F electrodes. This is in contrast to the F/I/F system which has different resistances for $P$ and $AP$ configurations, $R_{FIF}^{P,AP}$. By defining $\alpha_{P,AP}=R_{FIF}^{P,AP}/R_{FIS}$, the total resistance of the F/I/F/I/S can be expressed as
\begin{equation}
R_T=R_{FIF}+R_{FIS}=(\alpha_{P,AP}+1)R_{FIS}.
\end{equation}
With this notation, the Fano factor is given by
\begin{equation}
F_{P,AP}=(1+\alpha_{P,AP}^2)/(1+\alpha_{P,AP})^2,
\end{equation}
while the expression for TMR is
\begin{equation}
TMR=(\alpha_{AP}-\alpha_{P})/(1+\alpha_{P}).
\end{equation}
By using measured Fano factors in Fig.~\ref{fig7} averaged over bias with $F_P=0.94\pm0.1$ and $F_{AP}=0.58\pm0.1$, we obtain F/I/F/I/S TMR of about 40\% which is consistent with the TMR measured from the corresponding P and AP conductance, thus further corroborating tunneling without pinholes in our junctions. 

\section*{APPENDIX C: PERIODICITY OF ABOVE GAP CONDUCTANCE ANOMALIES}

In this section we present the conductance anomalies observed above the gap on F1/I/F2/I/S junctions, shown in Fig. \ref{fig5} in the main text, for a broader bias range. This results are similar to those found in Ref.~\onlinecite{Villegas2012} (Ref.~\onlinecite{Villegas2012} in the main text). The apparent lack of periodicity in Fig. \ref{fig5} in the main text is due to the limited bias range selected to show how the amplitude of one specific conductance anomaly varies with the four possible \textbf{M}-alignments.

\begin{figure}[H]
\begin{center}
\includegraphics[width=1\linewidth]{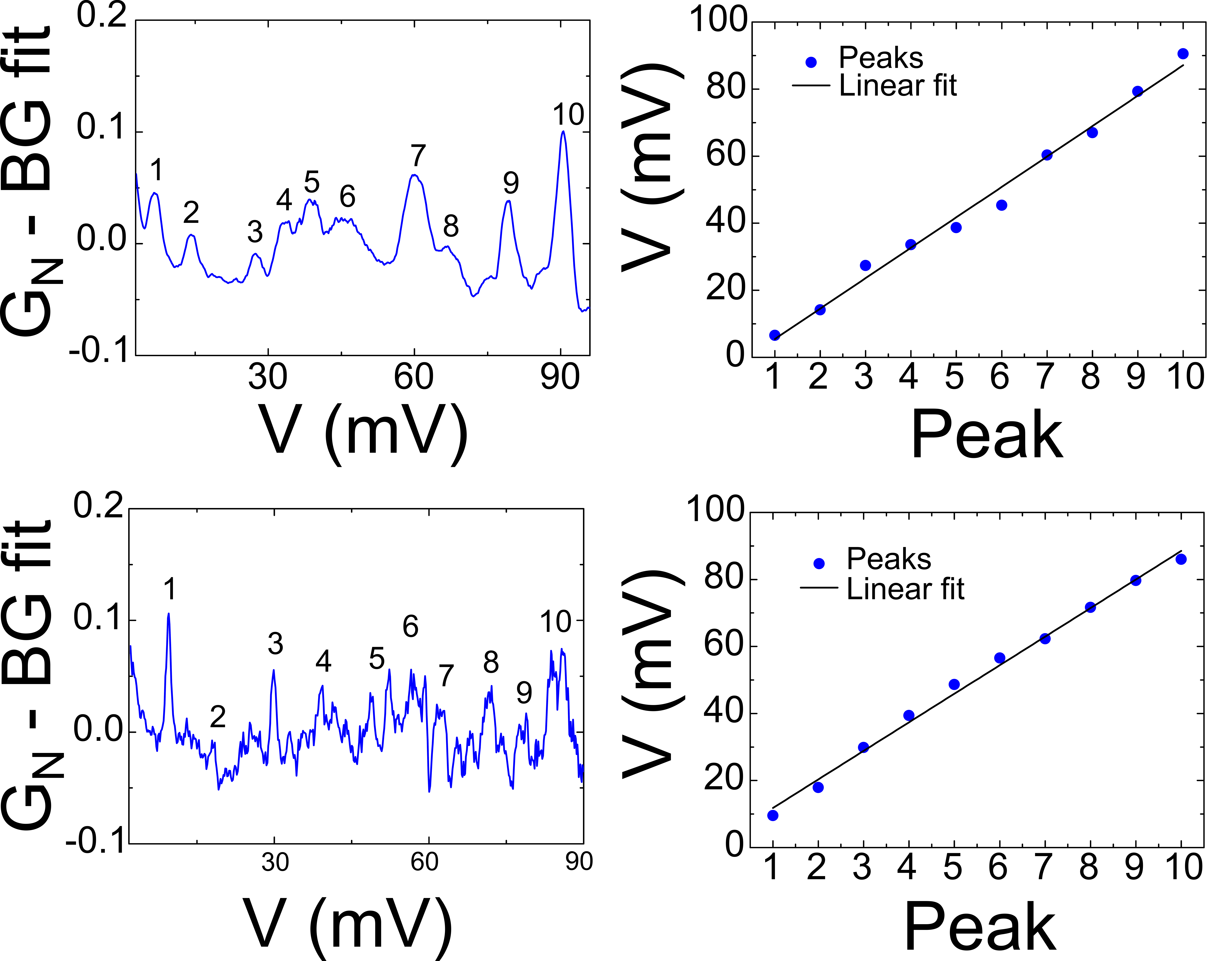}
\caption{Periodicity of  the above gap conductance anomalies for two different F1/I/F2/I/S junctions in the P state at 0.3 K. Left: normalized IV curves from above gap to $V\sim90$ mV, where a background fit (BG fit) has been subtracted from each $G_N(V)$ curve. The selected conductance anomalies (peaks) are marked with numbers. Right: Peak number versus bias for the two $G_N(V)$ curves, with a linear fit that shows how they fit to a periodic pattern. The lower panels are obtained from the same $G_N(V)$ curve shown in Fig. \ref{fig5}(a) of the manuscript. Only the bias positions of clearest peaks have been counted.}
\label{fig8}
\end{center}
\end{figure}

\section*{APPENDIX D: F/I/S MODEL IN THE PRESENCE OF SPIN-ORBIT COUPLING AND EXTERNAL MAGNETIC FIELD}
To model the F/I/S junction with interfacial spin-orbit coupling (SOC), where the F (S) region is semi-infinite at $z<0$ ($z>0$), we use a generalized BTK formalism~\onlinecite{Hogl2015} and solve the Bogoliubov-de Gennes equation~\onlinecite{deGennes1989} for quasiparticle states $\Psi(\mathbf{r})$ with energy $E$,
\begin{eqnarray}
\begin{pmatrix} \hat{H}_e & \hat{\Delta} \\ \hat{\Delta}^\dagger & \hat{H}_h  \end{pmatrix}\Psi(\mathbf{r})
&=& E \Psi(\mathbf{r}),
\label{eq:10}
\end{eqnarray}
where the single-particle Hamiltonian for electrons 
is $\hat{H}_e=-(\hbar^2/2) \boldsymbol{\nabla}\left[1/m(z)\right]\boldsymbol{\nabla}-\mu(z)- (\Delta_{xc}/2) 
\theta(-z) \mathbf{m}\cdot\boldsymbol{\hat{\sigma}}+ (V_0 d+\mathbf w \cdot \boldsymbol{\hat{\sigma}})\delta(z)$ 
and  for holes $\hat{H}_h=-\hat{\sigma}_y\hat{H}_e^*\hat{\sigma}_y$. 
They contain the effective mass $m(z)$, the chemical potential $\mu(z)$, and the exchange spin splitting $\Delta_{xc}$. The unit magnetization 
vector in the xz-plane is $\mathbf{m}=\left(\sin\theta , 0, \cos\theta\right)$ and 
$\boldsymbol{\hat{\sigma}}$ are Pauli matrices. The s-wave superconductor V is modeled by the pair potential 
$\hat{\Delta}= \Delta\theta(z) \mathbb{1}_{2\times2}$ with the isotropic gap $\Delta$.
At $z=0$ we assume a flat interface at which we account for the insulating MgO layer by including a delta-like potential barrier $V_0 d\delta(z)$, with effective height $V_0$ and width $d$. Since the built-in electric field due to structure inversion asymmetry inducing Rashba SOC is largest at interfaces, the Rashba SOC field~\onlinecite{Zutic2004,Fabian2007}
\begin{eqnarray}
\mathbf{w}_R=(\alpha k_y,-\alpha k_x),
\label{eq:2}
\end{eqnarray}
is also considered to be delta-like. Due to the conservation of in-plane wave vector $\mathbf{k}_{||}$, we can write 
$\Psi_{\sigma}(\mathbf{r})=\Psi_{\sigma}(z) e^{i\mathbf{k_{||}}\mathbf{r_{||}}}$. We find the solution in the F layer for incoming electrons with spin $\sigma$
\begin{eqnarray}
\Psi^{F}_{\sigma}&=&\frac{1}{\sqrt{k^e_{\sigma}}} e^{i k^e_{\sigma} z} \chi^e_{\sigma}+r^e_{\sigma, \sigma} 
e^{-i k^e_{\sigma} z} \chi^e_{\sigma}+ r^e_{\sigma, -\sigma} e^{-i k^e_{-\sigma} z} \chi^e_{-\sigma}  \nonumber \\
& & + r^h_{\sigma, -\sigma} e^{i k^h_{-\sigma} z} \chi^h_{-\sigma} +r^h_{\sigma, \sigma} e^{i k^h_{\sigma} z} 
\chi^h_{\sigma},
\label{eq:11}
\end{eqnarray}
with electron-like $\chi^e_{\sigma}=\left(\chi_{\sigma},0\right)^T$ and hole-like $\chi^h_{\sigma}=
\left(0,\chi_{-\sigma}\right)^T$ spinors, 
where
\begin{eqnarray}
\chi_\sigma^T=
\left(\sigma\sqrt{1+\sigma \cos\theta},
\sqrt{1-\sigma \cos\theta}\right)/\sqrt{2}
\label{eq:12}
\end{eqnarray}
and $\sigma=1(-1)$ corresponds to the spin parallel (antiparallel) to $\mathbf{\hat{m}}$. 
The scattering coefficients are: specular reflection $r^e_{\sigma, \sigma}$, specular reflection with spin flip $r^e_{\sigma, -\sigma}$, conventional Andreev reflection $r^h_{\sigma, -\sigma}$, and Andreev reflection with spin flip $r^h_{\sigma, \sigma}$.
The electron-like (hole-like) wave vectors in the F region are $k^{e(h)}_{\sigma}=
\sqrt{k_F^2+2m_{F}/\hbar^2\left[(-)E+ \sigma\Delta_{xc}/2\right]-k^2_{||}}$. 
The superconducting scattering states are
\begin{eqnarray} \nonumber
\Psi^{S}_{\sigma}&=& t^e_{\sigma, \sigma} e^{i q^e z}
\begin{pmatrix}
u \\
0\\
v \\
0
\end{pmatrix}
+t^e_{\sigma, -\sigma} e^{i q^e z}
\begin{pmatrix}
0 \\
u \\
0 \\
v
\end{pmatrix}\\
&+&  t^h_{\sigma, \sigma} e^{-i q^h z}
\begin{pmatrix}
v \\
0 \\
u  \\
0
\end{pmatrix}
+  t^h_{\sigma, -\sigma} e^{-i q^h z} 
\begin{pmatrix}
0 \\
v \\
0 \\
u 
\end{pmatrix},
\label{eq:13}
\end{eqnarray}
with superconducting coherence factors $u^2=1-v^2= \left(1+ \sqrt{E^2-\Delta^2}/|E|\right)/2$. 
The states comprise scattering coefficients for electron-like (hole-like) transmission without spin flip $t^e_{\sigma, \sigma}$ 
($t^h_{\sigma, \sigma}$) and electron-like (hole-like) transmission with spin flip $t^e_{\sigma, -\sigma}$ 
($t^h_{\sigma, -\sigma}$). The wave vectors are given by $\nolinebreak{q^{e(h)}=\sqrt{q_F^2+(-)2m_{S}/\hbar^2\sqrt{E^2-\Delta^2}-k^2_{||}}}$. 
Applying charge current conservation we compute the differential conductance at zero temperature
\begin{eqnarray}
G(V,\theta)=  \frac{e^2A}{(2\pi)^2h}\sum_{\sigma} \int{ d^2 \mathbf k_{\|}  \left[1+R^h_\sigma(-eV)-R^e_\sigma(eV)\right]}.
\label{eq:14}
\end{eqnarray}
Here the probability amplitudes in the F region,
$\nolinebreak{R^{e(h)}_{\sigma}(E,\mathbf{k_{\|}})=
	\mathrm{Re}\left( k^{e(h)}_{\sigma} 
	\left|r^{e(h)}_{\sigma,\sigma}\right|^2+k^{e(h)}_{-\sigma} \left|r^{e(h)}_{\sigma,-\sigma}\right|^2 \right)}$,
contain the scattering coefficients for specular and Andreev reflection with and without spin flip,  $A$ is the interfacial area, and we use Andreev approximation $k^e_\sigma=k^h_\sigma$.~\onlinecite{Blonder1982}

To describe the Fe/MgO/V junction we use the Fermi wave vectors $k_F=q_F=0.805 \times10^8$ cm$^{-1}$ for Fe and V, respectively. The effective 
masses are $m_{Fe}=m_V=m_0$ with $m_0$ the free electron mass. The spin polarization of Fe is given by $P=\left(\Delta_{xc}/2\right)/\mu_{Fe}=0.7$, where $\mu_{Fe}=\hbar^2k_F^2/(2m_{Fe})$. The  gap for V is $\Delta=0.8$ meV. The choice of the effective width and height of the epitaxially-grown MgO insulating barrier, $d=1.7$ nm and $V_0$ is guided by our low-bias experimental measurements. 
The SOC parameter $\alpha$ is adjusted to fit the experiments. The parameters are further discussed in Section E.
From the boundary conditions ensuring probability conservation 
 \begin{eqnarray}
\left.\Psi^{F}_{\sigma}\right|_{z=0^-}&=&\left.\Psi^{S}_{\sigma}\right|_{z=0^+},
\label{eq:15}
\end{eqnarray}
\begin{eqnarray}
&&\frac{\hbar^2}{2m_{S}} \frac{d}{dz} \eta  \left.\Psi^{S}_{\sigma}\right|_{z=0^+}=
\begin{pmatrix} \mathbf w \cdot \boldsymbol{\hat{\sigma}} & 0\\ 0 & -\mathbf w \cdot \boldsymbol{\hat{\sigma}}
\end{pmatrix} \left.\Psi^{F}_{\sigma}\right|_{z=0^-}  \nonumber\\
&& +\left(\frac{\hbar^2}{2m_{F}} \frac{d}{dz} +V_0d \right) \eta \left.\Psi^{F}_{\sigma}\right|_{z=0^-}, 
\label{eq:16}
\end{eqnarray}
with
\begin{eqnarray}
\eta &=& \begin{pmatrix} 
\mathbb{1}_{2\times2}&0\\
0&-\mathbb{1}_{2\times2}
\end{pmatrix},
\label{eq:17}
\end{eqnarray}
scattering coefficients are obtained numerically to give the resulting conductance by performing the integration in Eq.~(\ref{eq:14}). 
For a low bias $|eV|< \Delta$ quasiparticle transmission is prohibited and we get from probability current conservation $R^e_\sigma(eV)=1-R^h_\sigma(eV)$, which leads to
\begin{eqnarray}
G(V,\theta)=  \frac{e^2A}{(2\pi)^2h}\sum_{\sigma} \int{ d^2 \mathbf k_{\|}  \left[2R^h_\sigma(eV)\right]}. 
\label{eq:18}
\end{eqnarray}
Thus, zero bias conductance depends only on the probability amplitude of Andreev reflection. 
\begin{figure*}[]
	\begin{center}
	\includegraphics[width=2\columnwidth]{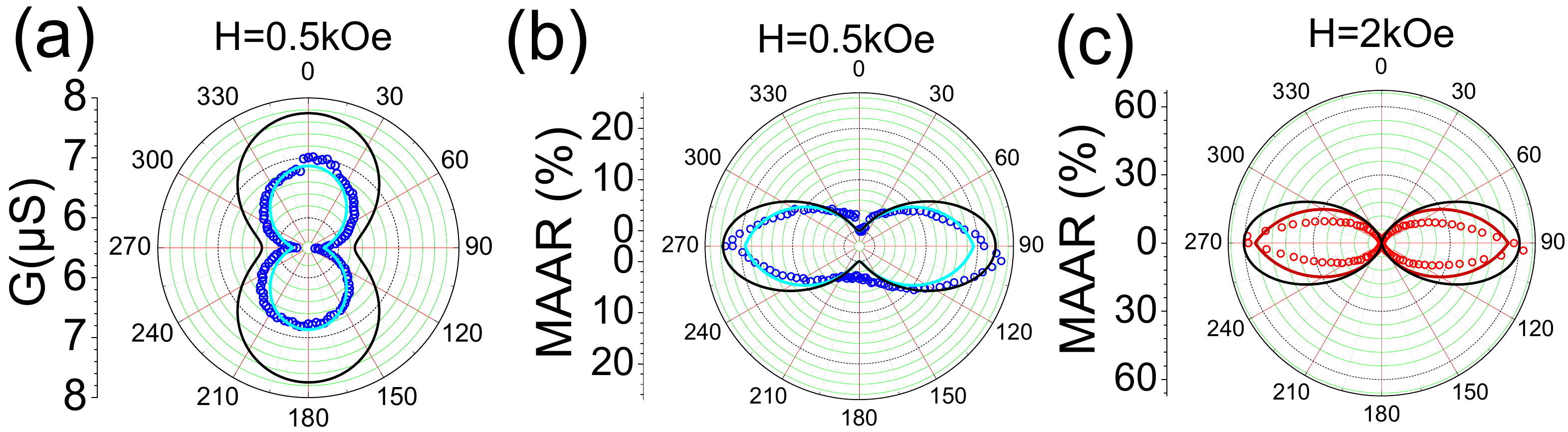}
			\caption{(a) Out-of-plane $G(V=0,\theta)$ 
			anisotropy measured at $T=0.3$ K and 
			$H=0.5$ kOe (blue dots) compared to: (i) fits from the extended BTK model 
			with $H=0$, $V_0=0.3$ eV, and $\alpha=3.0$ eV\AA$^2$ (black line) and (ii) the phenomenological model including 
			magnetic field effects with parameters $G_0=6.01\,\mathrm{\mu S}$, $A=-0.0415 \,\mathrm{\mu S}$, 
			and $B=-0.224 \,\mathrm{\mu S}$ (blue line). (b) The same for the of out-of-plane 
			MAAR($\theta$) at $H=0.5$ kOe. (c) Out-of-plane MAAR measured at $T=0.3$ K and $H=2$ kOe 
			(red dots) compared to: (i) fits from the extended BTK model with $H=0$, $V_0=0.3$ eV, and $\alpha=5.5$ eV\AA$^2$ 
			(black line) and (ii) the phenomenological model including magnetic field effects with $G_0$; $A$, 
			and $B$ as specified above (red line).}
		\label{fig9}
	\end{center}
\end{figure*}

The calculated angular dependence of $G$ and  MAAR are presented in Fig. \ref{fig9}. 
To get more insight into the physical mechanisms behind the angular dependence of $G$  and MAAR we consider a simple phenomenological model which was developed earlier for TAMR~\onlinecite{Fabian2007,Matos-Abiague2009} and also applied to MAAR.~\onlinecite{Hogl2015} The model is based on general 
symmetry arguments. It identifies two preferential directions in the system for given $\text{\textbf{k}}_\|$, namely $\mathbf{m}$ and $\mathbf{w}(\text{\textbf{k}}_\|)$. Therefore, a scalar quantity as the Andreev reflection probability can be expanded in powers of  $\mathbf{m}\cdot\mathbf{w}(\text{\textbf{k}}_\|)$. Up to second order in SOC we get for the conductance
\begin{eqnarray}
&G(V,\theta)=  \frac{e^2A}{(2\pi)^2h}\sum_{\sigma} \int{ d^2 \mathbf k_{\|}  2\left[R^{h,(0)}_\sigma(eV) \right.}& \label{eq:19} \\ \nonumber 
&\left.+R^{h,(1)}_\sigma(eV)\left[\mathbf{m}\cdot\mathbf{w}(\text{\textbf{k}}_\|)\right]       
+R^{h,(2)}_\sigma(eV)\left[\mathbf{m}\cdot\mathbf{w}(\text{\textbf{k}}_\|)\right]^2\right].& 
\end{eqnarray}
The linear term vanishes after integration due to $\mathbf{w}(\text{\textbf{k}}_\|)=-\mathbf{w}(-\text{\textbf{k}}_\|)$ 
and with Eq.~(\ref{eq:2}) we obtain 
\begin{eqnarray}
G(V,\theta)=G_0+g_2\alpha^2(1-\mathrm{cos}2\theta),
\label{eq:20}
\end{eqnarray}
which contains the SOC independent conductance 
$G_0=e^2A/(2\pi)^2/h\sum_{\sigma} \int{ d^2 \mathbf k_{\|}  2R^{h,(0)}_\sigma(eV)}$ and the SOC dependent  part with the expansion coefficient 
$g_2=e^2A/(2\pi)^2/h\sum_{\sigma} \int{ d^2 \mathbf k_{\|} R^{h,(2)}_\sigma(eV)k_y^2}$. For the MAAR we get 
\begin{eqnarray}
\mathrm{MAAR}(\theta)=\frac{g_2\alpha^2(\mathrm{cos}2\theta-1)}{G_0-g_2\alpha^2
	(\mathrm{cos}2\theta-1)}.
\label{eq:4}
\end{eqnarray}

To obtain the theoretical MAAR($\theta$) we choose the SOC value to match the experimental value at $\theta = 270^\circ$, by definition MAAR($0^\circ$)=0.  
In Fig.~\ref{fig9}(b) we observe a clear discrepancy between this MAAR($\theta$) fit (black line) and experiment at $H=0.5$ kOe (blue circles). The corresponding theoretical G in Fig.~\ref{fig9}(a) (black line), which was not fit, yields an even weaker agreement. At higher fields for $H=2$ kOe, 
as can be seen in Fig.~\ref{fig9}(c), the deviation from the measured MAAR($\theta$) is more pronounced. 
Since the MAAR amplitude grows with increasing $H$ we need two different values of the SOC parameter to match the magnitude in  Figs.~\ref{fig9}(b) and (c).
 
To take into account the influence of external magnetic field, required to rotate the magnetization in the $x-z$ plane, we recognize that there are two related effects: (i) orbital effects on the charge carriers and (ii) the suppression of superconductivity due to creation of vortices. 

We first discuss orbital effects. 
We assumed that the radius of the cyclotron orbits, on which charge carriers are forced by a magnetic field perpendicular to their propagation direction, is much larger than the width of the MgO tunnel barrier so that we can neglect effects from the external magnetic field. 
Since the influence of orbital effects from the external magnetic field is expected to be small, we include the magnetic field perturbatively to the phenomenological model for out-of-plane rotation of the magnetic field in a similar way as it was proposed for in-plane field.~\onlinecite{Wimmer2009}
A magnetic field is introduced to the model Hamiltonian using minimal coupling 
$\mathbf{p}=-i\hbar \boldsymbol{\nabla}\rightarrow \boldsymbol{\pi}=-i\hbar \boldsymbol{\nabla}+e\mathbf{A}$. 
The magnetic flux density is given by $\mathbf{B}=B\mathbf{m}$ so we can choose the gauge 
$\mathbf{A}=(-yB\mathrm{cos}\theta,-zB\mathrm{sin}\theta,0)$ for the vector potential $\mathbf{A}$ and relate 
it to the magnetic field by $\mathbf{B}=\mu_0\mathbf{H}$ with the vacuum permeability $\mu_0$. 
With the substitution for the momentum above, the kinetic energy and Rashba SOC in the single-particle Hamiltonian 
$\hat{H}_e$ are modified as 
$\hat{H}_{kin}=1/2 \boldsymbol{\pi}\left[1/m(z)\right]\boldsymbol{\pi}$ and 
$\hat{H}_R=\alpha/\hbar[(\pi_y,-\pi_x,0) \cdot \boldsymbol{\hat{\sigma}}]\delta(z)$, respectively. 
Instead of numerically solving this problem we want to study the underlying physical behavior.
Our strategy is to again expand the Andreev reflection probability in powers of magnetization direction and SOC field. This is still possible, however, the expansion coefficients and SOC field are now $H$-dependent. 
The field dependent quantities $R^{h,(n)}_{\sigma,H}(eV)$  
and $\mathbf{w}_H(\text{\textbf{k}}_\|)$ are approximated by the previous independent ones valid up to  linear order in $H$. We first look at the kinetic energy term. 
When $H=0$, the Andreev reflection is largest at $\mathbf{k}_\|=0$ since for finite $\mathbf{k}_\|$ 
a part of the total kinetic energy of an incoming electron is in the parallel direction to the interface which effectively increases the barrier height. When $H\ne0$ the maximum Andreev reflection is shifted to an in-plane wave vector $\mathbf{k}_{\|,0}$ which fulfills 
$\left<\left[k_{x,0}-e\mu_0Hy/\hbar\,\mathrm{cos}\theta \right]^2\right>=0$ and 
$\left<\left[k_{y,0}-e\mu_0Hz/\hbar\,\mathrm{sin}\theta \right]^2\right>=0$, performing a quantum mechanical average  $\left<...\right>$. 
Thus  the electrons feel effectively the smallest barrier for  $\mathbf{k}_{\|,0}=\left[b_1H\mathrm{cos}\theta,b_2H\mathrm{sin}\theta,0\right]$, 
where $b_1$ and $b_2$ are constants that depend on $\left<y\right>$ and $\left<y^2\right>$ or $\left<z\right>$ 
and $\left<z^2\right>$, respectively. Thus, we approximate 
$R^{h,(n)}_{\sigma,H}(\mathbf k_\|)\approx R^{h,(n)}_\sigma\left(\sqrt{(k_x-k_{x,0})^2+(k_y-k_{y,0})^2}\right)$. 
This shift can be related to the Lorentz force which sends the charge carriers on helicoids depending on the orientation of magnetization. Higher order $H$-effects on the Andreev reflection amplitude are neglected. The spin-orbit field experiences also a momentum shift 
$\mathbf{w}_H(\mathbf{k}_\|)\approx \mathbf{w}(k_x-b_3H\mathrm{cos}\theta,k_y-b_4H\mathrm{sin}\theta)$ with the coefficients $b_3$ and $b_4$, respectively depending on $\left<y\right>$ and $\left<z\right>$ because momentum appears linearly in the SOC field. 

In the presence of an out-of-plane magnetic field, the Hamiltonian is not anymore translationally invariant in the y-direction. We assume that we can treat those terms as small perturbations and compute the conductance from 
$G(V,\theta)=  e^2A/(2\pi)^2/h\sum_{\sigma} \int{ d^2 \mathbf k_{\|}  2R^{h}_\sigma(eV)}$ by using the expansion of the Andreev reflection probability in powers of SOC with $H$-dependent expansion coefficients. The SOC independent term is the same upon integration as when $H=0$. 
Conductance corrections to the second order term are neglected since they are quadratic in $H$. From the first order term, vanishing for $H=0$,  we get an additional contribution to the conductance due to the interplay of SOC and magnetic field
\begin{eqnarray}
G_{SOC-H}&=&\frac{e^2A}{(2\pi)^2h}\sum_{\sigma} \int{ d^2 \mathbf k_{\|}  2R^{h,(1)}_{\sigma,H}(eV)\left[\mathbf{m}\cdot 
\mathbf{w}(\mathbf{k}_\|)\right]}\nonumber \\
&=&g_1\alpha H(1-\mathrm{cos}2\theta), 
\label{eq:21}
\end{eqnarray}
which is linear in $H$ and contains the coefficient $g_1=e^2A/(2\pi)^2/h\sum_{\sigma} \int{ d^2 \mathbf k_{\|}'2R^{h,(1)}_{\sigma}(k_{\|}')(b_2-b_4)}$ 
with $\mathbf k_\|'=\mathbf k_\|-\mathbf k_{\|,0}$. So orbital effects in the presence of a Rashba SOC field induce an angular dependence which is of 
the same form as the one from the second order SOC term caused by the interplay of SOC and {\bf M} without considering the external magnetic field. The purely orbital contribution to the conductance, which is present for out-of-plane magnetic field even without SOC, is neglected because it is of higher order in $H$. The conductance which depends also on the $H$-field is 
\begin{eqnarray}
G(V,\theta,H)&=&G_0+(g_1\alpha H+g_2\alpha^2)(1-\mathrm{cos}2\theta)\nonumber \\
&=& G_0+G_\alpha (1-\mathrm{cos}2\theta).
\label{eq:22}
\end{eqnarray}
Considering only this $H$-correction is not enough since it does not cover the $G(\theta=0^\circ)$  increase [see Fig.~\ref{fig4}(a), main text] and the modified shape of the angular dependence.

We next consider a suppression of superconductivity in V as a type-II superconductor, due to vortex formation with magnetic field.
In our thin superconducting films, the measured critical field for a complete suppression of superconductivity is anisotropic:
$H_{C2,\perp}=3.5$ kOe ($H_{C2,\|}=12$ kOe) when $H$ is applied perpendicular (parallel)  to the film. 
The angular dependence of the critical field can be described by Tinkham's formula~~\onlinecite{Tinkham1963}
	\begin{eqnarray}
	 \frac{H_{C2}^2(\theta)}{H^2_{C2,\|}}\sin^2(\theta)+\frac{H_{C2}(\theta)}{H_{C2,\perp}}|\cos(\theta)|=1.
	 \label{eq:tinkham}
	\end{eqnarray}
	
We model the suppression of the superconductivity and the amount of normal conduction that is introduced by vortices by the ratio $h(\theta)=H/H_{C2}(\theta)$. We assume a linear increase (decrease) of the normal (superconducting) conductance contribution with  magnetic 
field.~\onlinecite{Miyoshi2005} The conductance becomes
\begin{eqnarray}
G(V,\theta,H)&=&h(\theta) G^N(V,\theta,H)\nonumber \\&+&[1-h(\theta)]G^S(V,\theta,H),
\label{eq:6}
\end{eqnarray}
with $G^j(V,\theta,H)=G_0^j+G_\alpha^j(1-\cos 2\theta)$, where $G_\alpha^j=A^jH/H_{C2,\perp}+B^j$ and $j=S,N$ 
are superconducting and normal  conductance. 
Solving Eq.~(\ref{eq:tinkham}) for $H_{C2}(\theta)$ gives
\begin{eqnarray}
H_{C2}(\theta)&=&-\frac{1}{2}\frac{H_{C2,\|}^2}{H_{C2,\perp}}\frac{\left|\cos\theta\right|}{\sin^2\theta}+\sqrt{\frac{1}{4}
\frac{H_{C2,\|}^4}{H_{C2,\perp}^2}\frac{\cos^2\theta}{\sin^4\theta}+\frac{H_{C2,\|}^2}{\sin^2\theta}}  \nonumber\\
&\mathrm{if}& \theta \neq 0^\circ, 180^\circ, \\
H_{C2}(\theta)&=&H_{C2,\perp}\nonumber \\
&\mathrm{if}& \theta = 0^\circ, 180^\circ. 
\end{eqnarray}

\section*{APPENDIX E: FITTING TO PHENOMENOLOGICAL MODEL}

From the phenomenological model we obtain the fitting 
\begin{eqnarray}
&&G(V,\theta,H)=\\ \label{eq3}
&=&h(\theta)\left[ G^N_0+(A^Nh(0^\circ)+B^N)(1-\cos 2\theta)\right]\nonumber  \\
&+&\left[1-h(\theta)\right]\left[ G^S_0+(A^Sh(0^\circ)+B^S)(1-\cos2\theta)\right]. \nonumber
\end{eqnarray}
We use the measurements from Fig.~\ref{fig10}(b) to assume
that $G^N=2G^S$ at $V=0$ and by omitting index S, Eq.~(\ref{eq3}) can be simplified as
\begin{equation}
G(\theta,H)=[1+h(\theta)]\left[G_0+(Ah(0^\circ)+B)(1-\cos 2\theta)\right], \label{eq4}
\end{equation}
where $G_0$, $A$, and $B$ need to be found. 
We apply the experimental values $H_{C2,\perp}=3.5$ kOe and $H_{C2,\|}=12$ kOe and determine $G_0$ from $G(\theta=0^\circ,H)$ at $H=0.5\,\mathrm{kOe}$ 
and $H=2\,\mathrm{kOe}$ as $G_0=5.92375\,\mathrm{\mu S}$ and $G_0=6.09\,\mathrm{\mu S}$, respectively. We use the average of both 
values $G_0=6.0069\,\mathrm{\mu S} \approx 6.01\,\mathrm{\mu S}$. Using $G(\theta=270^\circ,H=0.5\,\mathrm{kOe})$ and $G(\theta=270^\circ,H=2\,\mathrm{kOe})$ 
we get a set of two linear equations which we solve to obtain 
$A=-0.0415 \,\mathrm{\mu S}$ and $B=-0.224 \,\mathrm{\mu S}$. 
G$(\theta)$ and MAAR$(\theta)$ from this phenomenological model 
are shown in Figs.~\ref{fig9}(a)-(c).

\section*{APPENDIX F: F/I/S CONDUCTANCE AT FINITE TEMPERATURE}

\begin{figure}[h]
	\begin{center}
		\includegraphics[width=1\linewidth]{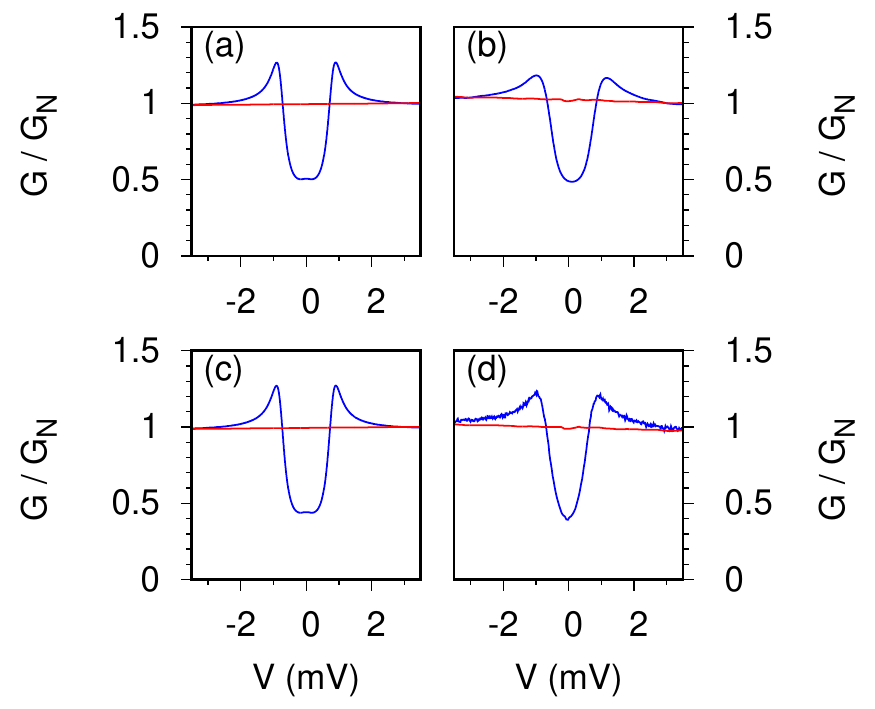}.        
		\caption{Calculated and measured F/I/S finite temperature conductance normalized  
				to $G_N=G(3 \mathrm{mV})$ for two different samples Fe/MgO/V at $H=0$. 
				(a) $G/G_N$ from extended BTK model in the superconducting  ($\Delta=0.8$ meV, blue line) 
				and normal state ($\Delta=0$, red line), $\alpha=4.6$ eV\AA$^2$, $Z=0.83$, $P=0.7$. 
				(b) Measured $G/G_N$ in the superconducting ($T=0.3$ K, blue line) 
				and normal state ($T=10$ K, red line).  
				The sample on which anisotropy measurements from Fig.~\ref{fig4} in the main text and Fig. \ref{fig9} in Appendix were taken. 
				(c) The same as in (a), but $\alpha=4.2$ eV\AA$^2$. 
				(d) The same as in (b) for the sample from measurements 
				shown in Fig. \ref{fig3} in the main text.}
		\label{fig10}
	\end{center}
\end{figure}

We fit the calculated conductance to typical experimental conductance spectra at $H=0$ 
with in-plane {\bf M}. 
We use the extended BTK model as described above and 
take into account finite temperature $T$.~\onlinecite{Hirai2003} In 
the fitting we adjust the barrier height $V_0$ and SOC $\alpha$ 
to match the characteristic properties of the conductance 
spectra.
\\
\\
A comparison between calculated and measured conductance spectra is shown in Fig.~\ref{fig10}. 
Our emphasis is not to obtain the best possible fits, but to recover the main features in the experimental data. For 
two measured samples, the calculated $G(V)$ in Figs.~\ref{fig4}(a) and (c) differ only in the choice of Rashba SOC, $\alpha$.
The ratio of the conductance peak 
near $V=0.8$ mV to the above gap conductance $G/G_N\approx 1.2$ suggests that the Fe/MgO/V junction is not in a strong tunneling limit. 
Consistent with the behavior of epitaxial MgO, we choose $V_0=0.3$ eV for the effective barrier height and use this value for all results 
from the extended BTK model at $T=0$ and finite 
$T$. This corresponds to $Z=V_0d\sqrt{m_{F}m_{S}}/(\hbar^2 \sqrt{k_Fq_F})=0.83$, a barrier parameter commonly used in the BTK model.~\onlinecite{Hogl2015,Blonder1982}


\begin{thebibliography}{99}


\bibitem{Zutic2004} I. \v{Z}uti\'c, J. Fabian and S. Das Sarma, Spintronics: Fundamentals and applications, \href{https://journals.aps.org/rmp/abstract/10.1103/RevModPhys.76.323}{Rev. Mod. Phys. \textbf{76}, 323 (2004).} 

\bibitem{Parkin2004} S. S. P. Parkin, C. Kaiser, A. Panchula, P. M. Rice, B. Hughes, M. Samant, and S.-H. Yang, Giant Tunnelling Magnetoresistance at Room Temperature with MgO (100) Tunnel Barriers, \href{https://www.nature.com/articles/nmat1256}{Nat. Mater. \textbf{3}, 862 (2004).}

\bibitem{Yuasa2004} S. Yuasa, T. Nagahama, A. Fukushima, Y. Suzuki and K. Ando, Giant Room Temperature Magneto-Resistance in Single-Crystal Fe/MgO/Fe Magnetic Tunnel Junctions, \href{https://www.nature.com/articles/nmat1257}{Nat. Mater. \href{3}, 868 (2004).}

\bibitem{Fabian2007} J. Fabian, A. Matos-Abiague, C. Ertler, P. Stano and I. \v{Z}uti\'c, Semiconductor spintronics, 
\href{http://www.physics.sk/aps/pubs/2007/aps-07-04/aps-07-04.pdf}{Acta Phys. Slov. \textbf{57}, 565 (2007).}

\bibitem{Singh2015} A. Singh, S. Voltan, K. Lahabi, and J. Aarts, Colossal Proximity Effect in a Superconducting Triplet Spin Valve Based on the Half-Metallic Ferromagnet CrO$_2$, \href{https://journals.aps.org/prx/abstract/10.1103/PhysRevX.5.021019}{Phys. Rev. X \textbf{5}, 021019 (2015).} 

\bibitem{Linder2015} J. Linder and J. W. A. Robinson, Superconducting Spintronics, \href{https://www.nature.com/articles/nphys3242}{Nat. Phys. 11, 307 (2015).}

\bibitem{Eschrig2011} M. Eschrig, Spin-Polarized Supercurrents for Spintronics, \href{https://physicstoday.scitation.org/doi/10.1063/1.3541944}{Phys. Today 64, 43 (2011).}

\bibitem{Guerrero2007} R. Guerrero, D. Herranz, F. G. Aliev, F. Greullet, C. Tiusan, M. Hehn, and F. Montaigne, High bias Voltage effect on Spin-Dependent Conductivity and Shot Noise in Carbon-Doped Fe(001)/MgO(001)/Fe(001) Magnetic Tunnel Junctions, \href{https://aip.scitation.org/doi/10.1063/1.2793619}{Appl. Phys. Lett.  \textbf{91}, 132504 (2007).}

\bibitem{Aliev2007} F. G. Aliev, R. Guerrero, D. Herranz, R. Villar, F. Greullet, C. Tiusan, and M. Hehn, Very Low $1/f$ Noise at Room Temperature in Fully Epitaxial Fe/MgO/Fe Magnetic Tunnel Junctions, \href{https://aip.scitation.org/doi/10.1063/1.2822812}{Appl. Phys. Lett. \textbf{91}, 232504 (2007).}

\bibitem{Buzdin2005} A. I. Buzdin, Proximity Effects in Superconductor-Ferromagnet Heterostructures, \href{https://journals.aps.org/rmp/abstract/10.1103/RevModPhys.77.935}{Rev. Mod.  Phys. \textbf{77}, 935 (2005).}

\bibitem{Bergeret2005} F. S. Bergeret, A. F. Volkov, and K. B. Efetov, Odd Triplet Superconductivity and Related Phenomena in Superconductor-Ferromagnet Structures, \href{https://journals.aps.org/rmp/abstract/10.1103/RevModPhys.77.1321}{Rev. Mod. Phys. \textbf{77}, 1321 (2005).}

\bibitem{Golubov2004}A. A. Golubov, M. Yu. Kupriyanov, and E. Il'ichev, The Current-Phase Relation in Josephson Junctions, \href{https://journals.aps.org/rmp/abstract/10.1103/RevModPhys.76.411}{Rev. Mod. Phys. \textbf{76}, 411 (2004).}

\bibitem{Banerjee2014} N. Banerjee, J. W. A. Robinson, and M. G. Blamire, Reversible Control of Spin-Polarized Supercurrents in Ferromagnetic Josephson junctions, \href{https://www.nature.com/articles/ncomms5771}{Nat. Commun. \textbf{5}, 4771 (2014).}

\bibitem{Baek2014} B. Baek, W. H. Rippard, S. P. Benz, S. E. Russek and P. D. Dresselhaus, Hybrid Superconducting Magnetic Memory Device Using Competing Order Parameters, \href{https://www.nature.com/articles/ncomms4888}{Nat. Commun. \textbf{5}, 3888 (2014).}

\bibitem{Gingrich2016} E. C. Gingrich, B. M. Niedzielski, J. A. Glick, Y. Wang, D. L. Miller, R. Loloee, W. P. Pratt Jr., and N. O. Birge, Controllable $0-\pi$ Josephson Junctions Containing a Ferromagnetic Spin Valve,  \href{https://www.nature.com/articles/nphys3681}{Nat. Phys. \textbf{12}, 564 (2016).}

\bibitem{Eschrig2019} M. Eschrig, Phase-Sensitive Interface and Proximity Effects in Superconducting Spintronics, in \emph{Spintronics Handbook: Spin Transport and Magnetism}, edited by E. Y. Tsymbal and I. \v{Z}uti\'c, 2nd ed. (CRC Press, Taylor \& Francis, Boca Rat\'on, FL, 2019).

\bibitem{Robinson2010} J. W. A. Robinson, J. D. S. Witt and M. G. Blamire, Controlled Injection of Spin-Triplet Supercurrents into a Strong Ferromagnet, \href{https://science.sciencemag.org/content/329/5987/59}{Science 329, \textbf{59} (2010).}

\bibitem{Keizer2006} R. S. Keizer, S. T. B. Goennenwein, T. M. Klapwijk, G. Miao, G. Xiao, and  A. A. Gupta, A Spin Triplet Supercurrent through the Half-Metallic Ferromagnet CrO$:2$, \href{https://www.nature.com/articles/nature04499}{Nature \textbf{439}, 825 (2006).}

\bibitem{Eschrig2003} M. Eschrig, J. Kopu, J. C. Cuevas and G. Schon, Theory of Half-Metal/Superconductor Heterostructures, \href{https://journals.aps.org/prl/abstract/10.1103/PhysRevLett.90.137003}{Phys. Rev. Lett. \textbf{90}, 137003 (2003).}

\bibitem{Gorkov2001} L. P. Gorkov and E. I. Rashba, Superconducting 2D System with Lifted Spin Degeneracy Mixed Singlet-Triplet State, \href{https://journals.aps.org/prl/abstract/10.1103/PhysRevLett.87.037004}{Phys. Rev. Lett. \textbf{87}, 037004 (2001).}

\bibitem{Bergeret2013} F. S. Bergeret and I. V. Tokatly, Singlet-Triplet Conversion and the Long-Range Proximity Effect in Superconductor-Ferromagnet Structures with Generic Spin Dependent Fields, \href{https://journals.aps.org/prl/abstract/10.1103/PhysRevLett.110.117003}{Phys. Rev. Lett. \textbf{110}, 117003 (2013).}

\bibitem{Hogl2015} P. H\"{o}gl, A. Matos-Abiague, I. \v{Z}uti\'c, and J. Fabian, Magnetoanisotropic Andreev reflection in ferromagnet-superconductor junctions, \href{https://journals.aps.org/prl/abstract/10.1103/PhysRevLett.115.116601}{Phys. Rev. Lett., \textbf{115}, 116601 (2015).} 

\bibitem{Cascales2012} J. P. Cascales, D. Herranz, F. G. Aliev, T. Szczepanski, V. K. Dugaev, J. Barnas, A. Duluard, M. Hehn, and C. Tiusan, Controlling Shot Noise in Double-Barrier Magnetic Tunnel Junctions, \href{https://journals.aps.org/prl/abstract/10.1103/PhysRevLett.109.066601}{Phys. Rev. Lett. \textbf{109}, 066601 (2012).} 

\bibitem{Zutic2019} I. \v{Z}uti\'c, A. Matos-Abiague, B. Scharf, H. Dery, and K. D. Belashchenko, Proximitized Materials, \href{https://www.sciencedirect.com/science/article/pii/S1369702118301111}{Materials Today \textbf{22}, 85 (2019).}

\bibitem{Tsymbal2011} K. D. Belashchenko and E. Y. Tsymbal, \emph{Handbook of Spin Transport and Magnetism}, edited by E. Y. Tsymbal and I. \v{Z}uti\'c (CRC Press, Boca Raton, FL, 2012). 

\bibitem{Blaha2001} P. Blaha, K. Schwarz, G. K. H. Madsen, D. Kvasnicka, and J. Luitz, WIEN2k, An Augmented Plane Wave + Local Orbitals Program for Calculating Crystal Properties \href{https://www.researchgate.net/publication/237132866_WIEN2k_An_Augmented_Plane_Wave_plus_Local_Orbitals_Program_for_Calculating_Crystal_Properties}{(TU Vienna, Vienna, 2001).}

\bibitem{Tiusan2007} C. Tiusan, M. Hehn, F. Montaigne, F. Greullet, S. Andrieu, and A. Schuhl, Spin Tunneling Phenomena in Single Crystal Magnetic Tunnel Junction Systems, \href{https://iopscience.iop.org/article/10.1088/0953-8984/19/16/165201/meta}{J. Phys.: Condens. Matter \textbf{19}, 165201 (2007).}

\bibitem{Butler2001} W. H. Butler, X.-G. Zhang, T. C. Schulthess, and J. M. MacLaren, Spin-Dependent Tunneling Conductance of Fe|MgO|Fe Sandwiches, \href{https://journals.aps.org/prb/abstract/10.1103/PhysRevB.63.054416}{Phys. Rev. B \textbf{63}, 054416 (2001).}

\bibitem{Martinez2018} I. Mart\'inez, C. Tiusan, M. Hehn, M. Chshiev, and F. G. Aliev, Symmetry Broken Spin Reorientation Transition in Epitaxial MgO/Fe/MgO Layers with Competing Anisotropies, \href{https://www.nature.com/articles/s41598-018-27720-7}{Sci. Rep. \textbf{8}, 9463 (2018).}

\bibitem{Blonder1982} G. E. Blonder, M. Tinkham, and T. M. Klapwijk, Transition from metallic to tunneling regimes in superconducting microconstrictions: 
Excess current, charge imbalance, and supercurrent conversion, \href{https://journals.aps.org/prb/abstract/10.1103/PhysRevB.25.4515}{Phys. Rev. B \textbf{25}, 4515 (1982).} 

\bibitem{Soulen1998} B. Soulen Jr., J. M. Byers, M. S. Osofsky, B. Nadgorny, T. Ambrose, S. F. Cheng, P. R. Broussard, C. T. Tanaka, J. Nowak, J. S. Moodera, A. Barry and J. M. D. Coey, Measuring the Spin Polarization of a Metal with a Superconducting Point Contact, \href{https://science.sciencemag.org/content/282/5386/85.long}{Science 282, \textbf{85} (1998).}

\bibitem{Parker2002} J. S. Parker, S. M. Watts, P. G. Ivanov, and P. Xiong, Spin Polarization of CrO$_2$ at and across an Artificial Barrier, \href{https://journals.aps.org/prl/abstract/10.1103/PhysRevLett.88.196601}{Phys. Rev. Lett. \textbf{88}, 196601 (2002).}

\bibitem{Nadgorny2012} B. E. Nadgorny, Point Contact Andreev Reflection Spectroscopy in \emph{Handbook of Spin Transport and Magnetism}, edited by E. Y. Tsymbal and I. \v{Z}uti\'c (CRC Press, Boca Raton, FL, 2012).

\bibitem{Ren2007} C. Ren, J. Trbovic, R. L. Kallaher, J. G. Braden, J. S. Parker, S. von Moln\'ar and P. Xiong, Measurement of the Spin Polarization of the Magnetic Semiconductor EuS with Zero-Field and Zeeman-Split Andreev Reflection Spectroscopy,\href{https://journals.aps.org/prb/abstract/10.1103/PhysRevB.75.205208}{ Phys. Rev. B \textbf{75}, 205208 (2007).}

\bibitem{Miyoshi2005} Y. Miyoshi,Y. Bugoslavsky, and L. F. Cohen, Andreev reflection spectroscopy of 
niobium point contacts in a magnetic field, \href{https://journals.aps.org/prb/abstract/10.1103/PhysRevB.72.012502}{Phys. Rev. B \textbf{72}, 012502 (2005)} 

\bibitem{Sangiao2011} S. Sangiao, J. M. De Teresa, M. R. Ibarra, I. Guillamon, H. Suderow, S. Vieira, and L. Morellon, Andreev Reflection under High Magnetic Fields in Ferromagnet-Superconductor Nanocontacts, \href{https://journals.aps.org/prb/abstract/10.1103/PhysRevB.84.233402}{Phys. Rev. B \textbf{84}, 233402 (2011).}

\bibitem{Yates2013} K. A. Yates, M.S. Anwar and J. Aarts , O. Conde , M. Eschrig , T. Lofwander  L.  F. Cohen, Andreev Spectroscopy of CrO$_2$ Thin Films on TiO$_2$ and Al$_2$O$_3$, \href{https://iopscience.iop.org/article/10.1209/0295-5075/103/67005/meta}{EPL \textbf{103}, 67005 (2013).}

\bibitem{Zutic1999} I. \v{Z}uti\'c and S. Das Sarma, Spin-Polarized transport and Andreev Reflection in Semiconductor/Superconductor Hybrid Structures, \href{https://journals.aps.org/prb/abstract/10.1103/PhysRevB.60.R16322}{Phys. Rev. B \textbf{60},16322(R) (1999).}


\bibitem{Vansteenkiste2014} A. Vansteenkiste, J. Leliaert, M. Dvornik, M. Helsen, F. Garcia-Sanchez, B. Van Waeyenberge, The Design and Verification of MuMax3. \href{https://aip.scitation.org/doi/full/10.1063/1.4899186}{AIP Advances \textbf{4}, 107133 (2014).}

\bibitem{Alekseevskii1976} N. E. Alekseevskii,  V. M. Sakosarenko,  K. Bl\"{u}thner, and H.‐J. K\"{o}hler, Superconducting Properties of Vanadium Films, \href{https://onlinelibrary.wiley.com/doi/abs/10.1002/pssa.2210340216}{Phys. Status Solidi A \textbf{34}, 541 (1976).}

\bibitem{Sekula1972} T. Sekula and R. H. Kernohan, Magnetic Properties of Superconducting Vanadium, \href{https://journals.aps.org/prb/abstract/10.1103/PhysRevB.5.904}{Phys. Rev. B \textbf{5}, 904 (1972).}

\bibitem{Tinkham1963} M. Tinkham, Effect of fluxoid quantization on transitions of superconducting films, 
\href{https://journals.aps.org/pr/abstract/10.1103/PhysRev.129.2413}{Phys. Rev. \textbf{129}, 2413 (1963)} 

\bibitem{Villegas2012}  C. Visani, Z. Sefrioui, J. Tornos, C. Leon, J. Briatico, M. Bibes, A. Barth\'el\'emy, J. Santamar\'ia, and E. Villegas, Equal-Spin Andreev Reflection and Long-Range Coherent Transport in high-Temperature Superconductor/Half-Metallic Ferromagnet Junctions, \href{https://www.nature.com/articles/nphys2318}{Nat. Phys. \textbf{8}, 539 (2012).} 

\bibitem{deJong1995} M. J. M. de Jong and C. W. J. Beenakker, Andreev Reflection in Ferromagnet-Superconductor Junctions, \href{https://journals.aps.org/prl/abstract/10.1103/PhysRevLett.74.1657}{Phys. Rev. Lett. \textbf{74}, 1657 (1995).}

\bibitem{Jeon2018} K.-R. Jeon, C. Ciccarelli, A. J. Ferguson, H. Kurebayashi, L. F. Cohen, X. Montiel, M. Eschrig, J. W. A. Robinson and M. G. Blamire, Enhanced Spin Pumping into Superconductors Provides Evidence for Superconducting Pure Spin Currents, \href{https://www.nature.com/articles/s41563-018-0058-9}{Nat. Mater. \textbf{17}, 499 (2018).}

\bibitem{Banerjee2018} N. Banerjee, J. A. Ouassou, Y. Zhu, N. A. Stelmashenko, J. Linder, and M. G. Blamire, Controlling the Superconducting Transition by Spin-Orbit Coupling, \href{https://journals.aps.org/prb/abstract/10.1103/PhysRevB.97.184521}{Phys. Rev. B \textbf{97}, 184521 (2018).}

\bibitem{Johnes2019} L. G. Johnes, N. Banerjee, and J. Linder, Magnetization Reorientation due to the Superconducting Transition in Heavy-Metal Heterostructures, \href{https://journals.aps.org/prb/abstract/10.1103/PhysRevB.99.134516}{Phys. Rev. B \textbf{99}, 134516 (2019).}

\bibitem{Blanter2000}
Ya. M. Blanter and M.  B\"{u}ttiker, Shot noise in mesoscopic conductors,
\href{https://arxiv.org/abs/cond-mat/9910158}{Phys. Rep. \textbf{336}, 1 (2000).} 

\bibitem{Szczepanski2013} T. Szczepa\ifmmode \acute{n}\else \'{n}\fi{}ski, V. K. Dugaev, J. Barna\ifmmode \acute{s}\else \'{s}\fi{}, 
J. P. Cascales and F. G. Aliev, Shot noise in magnetic double-barrier tunnel junctions, \href{https://journals.aps.org/prb/abstract/10.1103/PhysRevB.87.155406}{Phys. Rev. B \textbf{87}, 155406 (2013).} 

\bibitem{deGennes1989} P. G. De\hspace{0.1cm}Gennes, {\em Superconductivity of Metals and Alloys} (Addison-Wesley,  
Reading, MA 1989).


\bibitem{Matos-Abiague2009} A. Matos-Abiague and J. Fabian, Anisotropic tunneling magnetoresistance and 
tunneling anisotropic magnetoresistance: Spin-orbit coupling in magnetic tunnel junctions, \href{https://journals.aps.org/prb/abstract/10.1103/PhysRevB.79.155303}{Phys. Rev. B \textbf{79}, 155303 (2009).} 

\bibitem{Wimmer2009} M. Wimmer, M. Lobenhofer, J. Moser, A. Matos-Abiague, D. Schuh, W. Wegscheider, J. Fabian, K. Richter and D. Weiss, 
Orbital effects on tunneling anisotropic magnetoresistance in Fe/GaAs/Au junctions, \href{https://journals.aps.org/prb/abstract/10.1103/PhysRevB.80.121301}{Phys. Rev. B \textbf{80}, 121301 (2009).} 

\bibitem{Hirai2003} T. Hirai,  Y. Tanaka, N.  Yoshida, Y. Asano, J. Inoue, and S. Kashiwaya,
Temperature dependence of spin-polarized transport in ferromagnet/unconventional superconductor junctions, 
\href{https://journals.aps.org/prb/abstract/10.1103/PhysRevB.67.174501}{Phys. Rev. B \textbf{67}, 174501 (2003)} 


\end{thebibliography}
\end{document}